\documentclass[fleqn,10pt]{wlscirep}
\usepackage[utf8]{inputenc}
\usepackage[T1]{fontenc}
\usepackage{xcolor}
\usepackage[version=4]{mhchem} 
\usepackage{siunitx} 
\usepackage{soul}
\usepackage{pdfpages}
\usepackage{caption}
\usepackage{eso-pic} 
\usepackage{rotating}
\usepackage{makecell}
\usepackage{upquote}

\DeclareUnicodeCharacter{2032}{'}

\newcounter{extendeddatafig}

\newenvironment{extendeddatafigure}{
    \refstepcounter{extendeddatafig}
    \captionsetup{labelformat=empty}
    \begin{figure}[htbp]
    
}{
    \end{figure}
}

\newcounter{sifig}

\newenvironment{sifig}{
    \refstepcounter{sifig}
    \captionsetup{labelformat=empty}
    \begin{figure}[htbp]
    
}{
    \end{figure}
}

\title{Terahertz control of surface topology probed with subatomic resolution}

\author[1,+]{V. Jelic}
\author[1,+]{S. Adams}
\author[2,+]{D. Maldonado-Lopez}
\author[1,2]{I. A. Buliyaminu}
\author[1]{M. Hassan}
\author[1,2 *]{J. L. Mendoza-Cortes}
\author[1, *]{T. L. Cocker}

\affil[1]{Department of Physics and Astronomy, Michigan State University, East Lansing, MI 48824, USA}
\affil[2]{Department of Chemical Engineering and Materials Science, Michigan State University, East Lansing, MI 48824, USA}

\affil[+]{These authors contributed equally to this work.}
\affil[*]{jmendoza@msu.edu, cockerty@msu.edu}

\begin{abstract}

Light-induced phase transitions offer a method to dynamically modulate topological states in bulk complex materials\cite{basov2017towards, Torre2021, Ma2021, Bao2022}. Yet, next-generation devices demand nanoscale architectures with contact resistances near the quantum limit\cite{li2023approaching} and precise control over local electronic properties\cite{Zhai2024}. The layered material \ce{WTe2} has gained attention as a likely Weyl semimetal\cite{soluyanov2015type, Wang_observation_2016, Bruno_observ_2016, wu_observ_2016, SanchezBarriga2016, Feng_spin_2016, DiSante2017, Zhang2017RRL, li_evidence_2017, lin_visualizing_2017, Zhang_qpi_2017, Yuan_qpi_2018, Caputo_dynamics_2018, Das_electronicproperties_2019, sie_ultrafast_2019, Hein2020mode, guan_manipulating_2021, Drueke2021}, with topologically protected linear electronic band crossings hosting massless chiral fermions. Here, we demonstrate a topological phase transition facilitated by light-induced shear motion of a single atomic layer at the surface of bulk \ce{WTe2}, thereby opening the door to nanoscale device concepts. Ultrafast terahertz fields enhanced at the apex of an atomically sharp tip\cite{cocker_ultrafast_2013, Cocker2016,jelic2017ultrafast, peller2021quantitative, cocker_nanoscale_2021, Jelic_atomicTHzTDS_2024} resonantly couple to the key interlayer shear mode of \ce{WTe2} via a ferroelectric dipole at the interface\cite{fei2018ferroelectric, Yang2018, Ni2020, Xiao2020}, inducing a structural phase transition at the surface to a metastable state. Subatomically resolved differential imaging, combined with hybrid-level density functional theory, reveals a shift of 7~$\pm$~3 picometres in the top atomic plane. Tunnelling spectroscopy links electronic changes across the phase transition with the electron and hole pockets in the band structure, suggesting a reversible, light-induced annihilation of the topologically-protected Fermi arc surface states in the top atomic layer.
      
\end{abstract}

\begin{document}

\captionsetup[figure]{labelformat={empty}, justification=justified}

\flushbottom
\maketitle
\thispagestyle{fancy}


\section*{Introduction}

With technology reaching the limits of conventional design and fabrication, next-generation device concepts based on nanoscale architectures, terahertz clock rates, and complex material systems are becoming increasingly important. For example, light-induced phase transitions in complex materials promise a route to ultrafast switching of versatile electronic and photonic devices\cite{basov2017towards, Torre2021, Ma2021, Bao2022}. Materials with switchable, topologically non-trivial phases are especially intriguing due to their symmetry-protected states, which offer unique electronic properties.

Theoretical calculations\cite{soluyanov2015type, Wang_observation_2016, Feng_spin_2016, Bruno_observ_2016, DiSante2017, li_evidence_2017, Zhang2017RRL, lin_visualizing_2017, Zhang_qpi_2017, Yuan_qpi_2018, sie_ultrafast_2019, Das_electronicproperties_2019, Hein2020mode, kwon2020quasiparticle, guan_manipulating_2021, SanchezBarriga2016} and experimental studies with scanning tunnelling microscopy (STM)\cite{lin_visualizing_2017, Zhang_qpi_2017, Yuan_qpi_2018}, angle-resolved photoemission spectroscopy (ARPES)\cite{wu_observ_2016, DiSante2017, Caputo_dynamics_2018, Bruno_observ_2016, Wang_observation_2016, Feng_spin_2016, SanchezBarriga2016}, and transport measurements\cite{li_evidence_2017} suggest that the non-centrosymmetric T\textsubscript{d} phase of bulk \ce{WTe2} is a type-II Weyl semimetal at low temperature. This classification refers to the presence of three-dimensional (3D) linear electronic band crossings at the touching points of electron and hole pockets. The absence of inversion symmetry leads to pairs of Weyl points with opposite chirality, which are non-degenerate in momentum space and connected by topologically protected Fermi arc surface states\cite{soluyanov2015type}. It has been shown that strong terahertz fields can resonantly excite an interlayer shear mode\cite{sie_ultrafast_2019}, while mid-infrared and near-infrared pulses can non-resonantly drive the bulk lattice structure into a metastable state via the shear mode\cite{He2016softening, Hein2020mode, ji_manipulation_2021, guan_manipulating_2021, Drueke2021, Soranzio2022, sie_ultrafast_2019, Qi2022intralayer}. The light-induced structural phase transition can restore inversion symmetry, which leads to an annihilation of the Weyl points and, hence, a topological electronic phase transition from a Weyl semimetal to a trivial semimetal. 

In contrast to its bulk form, monolayer \ce{WTe2} is a quantum spin Hall insulator in its low-temperature ground state\cite{Tang2017}. Moreover, the stacking of few-layer \ce{WTe2} leads to a spontaneous out-of-plane electric polarization\cite{fei2018ferroelectric, Yang2018, Ni2020, Xiao2020}. The orientation of this ferroelectric dipole can be switched through a sliding motion of the top atomic layer, which can be induced by a vertical field on the scale of 0.1--0.2 V/nm (refs.\,[\citen{fei2018ferroelectric, Xiao2020}]). A similar shift of the top atomic layer of bulk \ce{WTe2} has also been activated by static doping with adsorbed potassium atoms\cite{Rossi2020}. In the present work, we show that terahertz fields enhanced at an atomically sharp tip apex and polarized normal to a bulk \ce{WTe2} surface couple to its ferroelectric dipole. When these terahertz near-fields exceed \qty{1}{V/nm}, they induce a localized structural phase transition of the topmost atomic layer, driving it from the low-temperature T\textsubscript{d} phase to a metastable state characterized by both a shear translation and an intralayer distortion. As a result, the electronic structure at the vacuum interface undergoes a phase transition, where the top layer can no longer host Fermi arc surface states. Density functional theory (DFT) calculations using hybrid exchange-correlation functionals are integral to this analysis, providing a comprehensive understanding of the ground and metastable states observed in experiments along with insight into the stability of the Weyl points.


\section*{Results}

\subsection*{Subsection 1: Shear motion induced by tip-enhanced terahertz fields}

\begin{figure}
    \centering
    \includegraphics[width=1\linewidth]{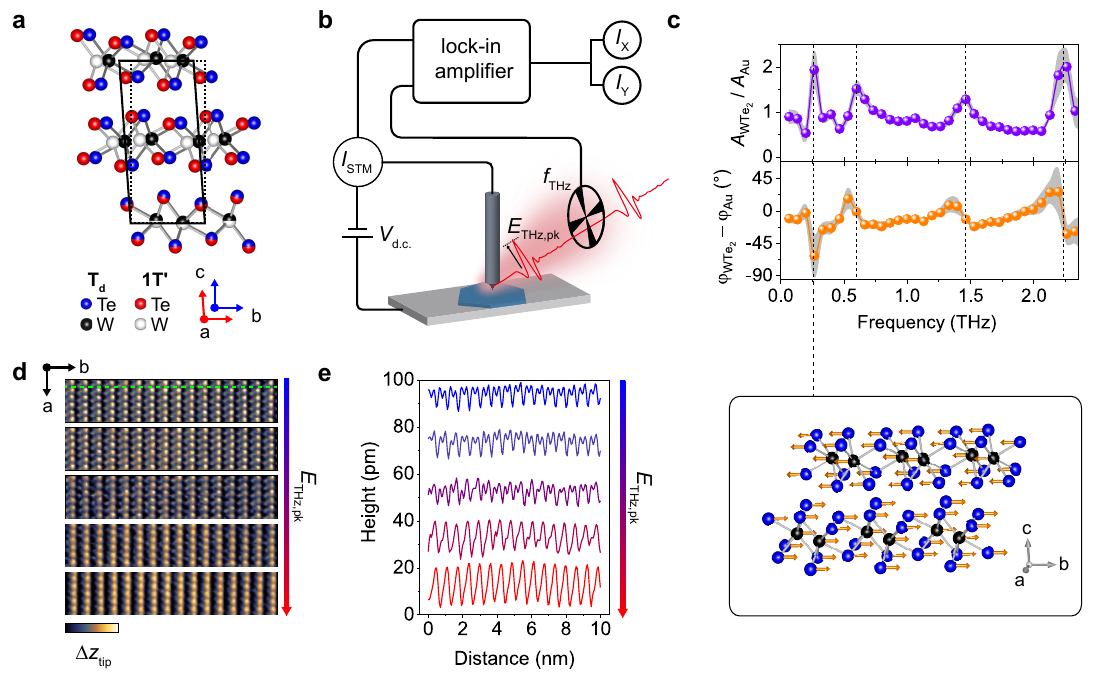}
    \caption{
        \textbf{Figure 1 $\mid$ Shear motion in \ce{WTe2} driven by tip-enhanced terahertz fields.} 
        \textbf{a}, Unit cells for the orthorhombic T\textsubscript{d} phase (blue and black spheres) and monoclinic 1T\textquotesingle{} phase (red and white spheres). The solid black line highlights the $\approx$\ang{4} tilt of the unit cell for 1T\textquotesingle{} with respect to T\textsubscript{d} (dashed black line).
        \textbf{b}, The terahertz-pulse-train incident on the scanning tunnelling microscopy (STM) tip is modulated at $f_\mathrm{THz}$ and the terahertz-pulse-induced contribution to the total tunnel current ($I_\mathrm{STM}$) is isolated with a lock-in amplifier, producing both in-phase ($I_\mathrm{X}$) and out-of-phase ($I_\mathrm{Y}$) components. A steady-state bias ($V_\mathrm{d.c.}$) continuously tunnels electrons at the tip apex, while a terahertz pulse with peak amplitude $E_\mathrm{THz,pk}$ incident on the tip (red curve), drives a phase transition at the tip apex and/or tunnels electrons between the tip and sample.
        \textbf{c}, Terahertz time-domain spectroscopy (THz-TDS) of the tunnel junction\cite{Jelic_atomicTHzTDS_2024} shows resonances at \qty{0.26}{THz}, \qty{0.60}{THz}, \qty{1.46}{THz} and \qty{2.24}{THz} (dashed black lines) in both the spectral amplitude (top, purple points) and phase (orange points). The THz-TDS shown here utilizes a reference waveform acquired on a gold surface\cite{Jelic_atomicTHzTDS_2024} (Extended Data Fig.\,\ref{fig:ext-validation}), $E_\mathrm{Au}(A_\mathrm{Au},\varphi_\mathrm{Au}) = \mathfrak{F}\{ E_\mathrm{Au}(t) \}$, with the same tip apex and incident terahertz alignment as the experiments on \ce{WTe2}. See Methods for parameters during waveform acquisition. The data is shown as mean values $\pm$ standard deviation of five individual scans. Bottom: schematic of the shear mode associated with a phase transition in the topmost layer of \ce{WTe2}, occurring at a frequency of \qty{0.26}{THz} (arrow lengths are exaggerated for clarity). 
        \textbf{d}, STM topography scans acquired at $V_\mathrm{d.c.} = \qty{10}{mV}$ and $I_\mathrm{d.c.} = \qty{100}{pA}$ with incident $E_\mathrm{THz,pk}$ at \qty{13}{V/cm} (top), \qty{18}{V/cm} (top middle), \qty{23}{V/cm} (middle), \qty{32}{V/cm} (bottom middle) and \qty{36}{V/cm} (bottom). Colourmap range [5,35] pm; image size $\qty{10}{nm} \times \qty{2}{nm}$; scan speed \qty{4.3}{nm/s}.
        \textbf{e}, Horizontal cross-sections of the respective topography scans in \textbf{d} (dashed green line). The cross-sections in \textbf{e} are vertically offset for clarity.
    }
    \label{fig:fig1}
\end{figure}

A single layer of \ce{WTe2} is comprised of tungsten atoms sandwiched between tellurium atoms (Fig.\,\ref{fig:fig1}a). In a bulk crystal, these layers are stacked through van der Waals bonding. Inter-metallic bonding between the tungsten atoms leads to zigzag chains in the $\vec{a}$-axis of the unit cell, which distorts the otherwise hexagonal structure and results in a corrugated surface\cite{kwon2020quasiparticle, Zhang_qpi_2017, Yuan_qpi_2018, lin_visualizing_2017, Chen_noncentrosym_2022}. The T\textsubscript{d} phase (blue and black atoms in Fig.\,\ref{fig:fig1}a), which is the low temperature ground state of \ce{WTe2}, has an orthorhombic unit cell with broken inversion symmetry that gives rise to two distinct surface terminations defined as \textquotesingle{}Surface A\textquotesingle{} and \textquotesingle{}Surface B\textquotesingle{} (see Methods for details). In 2015, T\textsubscript{d}-\ce{WTe2} was proposed as a candidate type-II Weyl semimetal\cite{soluyanov2015type}. In contrast, 1T\textquotesingle{}-\ce{WTe2} -- the lattice structure at high temperature\cite{Tao_Tdto1Tp_2020} or high pressure\cite{zhou_pressure_2016} -- is monoclinic and inversion symmetric (red and white atoms in Fig.\,\ref{fig:fig1}a), which prohibits Weyl points in the band structure. The transition from T\textsubscript{d}-\ce{WTe2} to 1T\textquotesingle{}-\ce{WTe2} is associated with a shift of the individual layers, resulting in a change of the angle between the unit cell vectors $\vec{b}$ and $\vec{c}$ from 90$^{\circ}$ to $\approx$94$^{\circ}$.

The experimental setup is depicted in Fig.\,\ref{fig:fig1}b. A train of single-cycle terahertz pulses is focused onto the tip of a scanning tunnelling microscope, which is in tunnel contact with the surface of a single-crystal \ce{WTe2} sample. The STM bias voltage, $V_\mathrm{d.c.}$, determines the difference between the Fermi levels of the tip and sample and thereby defines the energy range of electronic states contributing to  the tunnel current, $I_\mathrm{STM}$, from the \ce{WTe2} sample. Atomically resolved STM images can be recorded in constant-height mode or constant-current (topography) mode. The ultrafast terahertz pulses coupled to the STM tip experience a field enhancement of $>$10$^5$ at its apex\cite{jelic2017ultrafast, peller2021quantitative, Cocker2016, cocker_nanoscale_2021, Jelic_atomicTHzTDS_2024}. As a result, terahertz pulses with peak fields on the order of \qty{20}{V/cm} in free space generate transient fields of about \qty{1}{V/nm} across the tip–sample junction, driving a tunnel current measured via lock-in detection, with $I\textsubscript{X}$ and $I\textsubscript{Y}$ representing the in-phase and out-of-phase components, respectively (see Methods).

We perform atomic-scale terahertz time-domain spectroscopy\cite{Jelic_atomicTHzTDS_2024} to determine the dielectric response of the tip-sample junction at terahertz frequencies, as shown in Fig.\,\ref{fig:fig1}c (see Methods for details). The peaks at \qty{0.26}{THz}, \qty{0.60}{THz}, and \qty{2.24}{THz} in the experimental spectra agree with past literature\cite{guan_manipulating_2021, Hein2020mode, Soranzio2022, He2016softening, sie_ultrafast_2019, ji_manipulation_2021, Drueke2021, Qi2022intralayer} and our calculated phonon band structure at the $\Gamma$ point (Extended Data Fig.\,\ref{fig:ext-phononbandstructure} and Supplementary Table~\ref{tab:sitab-phonondispersion}). Meanwhile, the peak at \qty{1.46}{THz} may be spectrally bright due to the broad range of momentum vectors in the evanescent terahertz near-field at the tip apex\cite{cocker_nanoscale_2021}, which can excite phonons away from the $\Gamma$ point. The atomic motion for the \qty{0.26}{THz} mode is illustrated at the bottom of Fig.\,\ref{fig:fig1}c. This mode is associated with the phase transition from T\textsubscript{d} to 1T\textquotesingle{}\cite{guan_manipulating_2021, Hein2020mode, Soranzio2022, He2016softening, sie_ultrafast_2019, ji_manipulation_2021, Drueke2021, Qi2022intralayer}. Despite its primarily in-plane motion (and hence in-plane dipole), the \qty{0.26}{THz} shear mode is driven resonantly by the vertically-oriented terahertz near-fields, which couple to the out-of-plane ferroelectric dipole at the surface\cite{fei2018ferroelectric, Yang2018, Ni2020, Xiao2020}. 

Surprisingly, increasing the peak terahertz field in the STM junction (top-to-bottom in Fig.\,\ref{fig:fig1}d) modifies the \ce{WTe2} surface charge density observed in STM topography maps. This change is reversible, with the topography returning to its original state when the terahertz field is reduced. The top tellurium atoms of \ce{WTe2} are the main contributors to the surface charge density\cite{Chen_noncentrosym_2022}. Notably, along the $\vec{b}$-axis, both the upper and lower tellurium atoms of the corrugated surface are visible at low terahertz fields, whereas only the topmost atoms contribute substantially to the image at high fields. The evolution of the surface periodicity is further emphasized in the horizontal cross-sections along the $\vec{b}$-axis in Fig.\,\ref{fig:fig1}e, while Extended Data Fig.\,\ref{fig:ext-fourieranalysis} shows the corresponding evolution in the Fourier domain. Such a strong influence of terahertz fields on topography has not been previously observed, suggesting a long-lived excitation of the sample to a metastable state, rather than an ultrafast tunnel current induced by the terahertz pulses\cite{cocker_ultrafast_2013, cocker_nanoscale_2021, Cocker2016,jelic2017ultrafast, Jelic_atomicTHzTDS_2024}.

\subsection*{Subsection 2: Real-space differential imaging}

To explore the phase transition further, we modulate the terahertz pulse train at a frequency $f_\mathrm{THz}$ (Fig.\,\ref{fig:fig2}a, top). We identify three regimes from the temporal response of the tunnel current ($I_\mathrm{STM}$): For low field strengths (regime I), the terahertz pulse train does not affect the topography image or the tunnel current, and hence neither $I_\mathrm{X}$ nor $I_\mathrm{Y}$ is detected. For high field strengths (regime III), the terahertz pulse train induces a square-wave modulation of the tunnel current, producing a signal in $I_\mathrm{X}$ only ($I_\mathrm{Y} = 0$). This regime corresponds to lightwave-driven tunnelling, where the terahertz field acts as a quasi-static bias in the strong-field limit, as observed in previous THz-STM studies\cite{cocker_ultrafast_2013, cocker_nanoscale_2021, Cocker2016, jelic2017ultrafast, Jelic_atomicTHzTDS_2024} (see Methods for the terahertz field to voltage conversion). 

For intermediate field strengths (regime II), although the tunnel current is modulated at $f_\mathrm{THz}$, it only responds rapidly when the terahertz pulse-train is unblocked. When the terahertz pulses are blocked by the chopper, the tunnel current recovers exponentially on a timescale comparable to the chopping period (see Extended Data Fig.\,\ref{fig:ext-oscope} for further characterization). This behaviour of the tunnel current results from the terahertz-field-induced phase transition of T\textsubscript{d}-\ce{WTe2} to a metastable state, followed by a gradual return to the T\textsubscript{d} phase, as shown in Fig.\,\ref{fig:fig1}d,e. The metastable state resembles 1T\textquotesingle{}-\ce{WTe2} and will be discussed in detail in the following sections. Since the metastable state is long-lived, the sample does not decay to the T\textsubscript{d} ground state between individual terahertz pulses (spaced by 1 µs). Rather, the ground state is recovered during the time that the chopper blade blocks the terahertz pulse train. By comparison, in regime III, the excitation is strong enough that the ground state is not recovered when the pulse train is blocked and the sample remains in the metastable state. 

Regime II is observed for terahertz near-fields of $E_\mathrm{THz,pk} \approx$ \qty{1}{V/nm}, which are concentrated at the tip apex and decay radially outward. This is comparable to the static electric fields necessary for ferroelectric switching of few-layer \ce{WTe2}\cite{Yang2018, Xiao2020} and the off-resonant mid-infrared far-fields used to drive the phase transition in a bulk sample\cite{sie_ultrafast_2019}. However, regime II is only observed in the presence of a small STM bias voltage, which can be just a few mV, and therefore applies a field two to three orders of magnitude lower than the terahertz pulse across the \qty{1}{nm} tunnel gap. The signal measured by lock-in detection in regime II contains both an in-phase ($I_\mathrm{X}$) and out-of-phase ($I_\mathrm{Y}$) component due to the millisecond-scale recovery towards the ground state. This can be seen in terahertz scanning tunnelling spectroscopy, where sharp peaks emerge in both $I_\mathrm{X}$ and $I_\mathrm{Y}$ (Fig.\,\ref{fig:fig2}b), but only in the presence of a small but finite $V_\mathrm{d.c.}$. The d.c.\;bias is critical for regime II because the terahertz field drives the phase transition while the d.c.\;bias acts as read-out, with the detected signal corresponding to the difference in current between the T\textsubscript{d} phase and metastable phase for a given tip position and $V_\mathrm{d.c.}$. 

Since $I_\mathrm{X}$ also includes contributions from lightwave-driven tunnelling, only $I_\mathrm{Y}$ is exclusively associated with the difference measurement between the two phases. Figure~\ref{fig:fig2}c shows how $V_\mathrm{d.c.}$ affects atomically resolved $I_\mathrm{X}$ and $I_\mathrm{Y}$ images of a top-surface tellurium vacancy\cite{Chen_noncentrosym_2022}, with the terahertz field strength in regime II. For finite bias voltages of $V_\mathrm{d.c.} = \qty{\pm 5.5}{mV}$, the $I_\mathrm{Y}$ images exhibit similar spatial features and signal strengths but opposite polarity, whereas the $I_\mathrm{Y}$ signal is negligible for $V_\mathrm{d.c.} = 0$ V. Meanwhile, a purely lightwave-driven image is obtained in the $I_\mathrm{X}$ channel for $V_\mathrm{d.c.} = 0$ V, but for $V_\mathrm{d.c.} \neq 0$, $I_\mathrm{X}$ contains both lightwave-driven and phase-difference contributions.

By comparing images acquired over a larger area with multiple surface defects, we reveal a clear distinction between the two measurement modes. Lightwave-driven tunnelling (Fig.\,\ref{fig:fig2}d) reflects the local density of states (LDOS) up to the terahertz peak voltage, while the d.c.\;readout via $I\textsubscript{Y}$ (Fig.\,\ref{fig:fig2}e) probes the LDOS difference between the two phases up to $V_\mathrm{d.c.}$. The insets show two-dimensional fast Fourier transforms (FFTs) of the corresponding THz-STM scans. The features just to the left and right of the central peak can be associated with quasiparticle standing waves in the surface state, as observed through differential conductance imaging (Extended Data Fig.\,\ref{fig:ext-qpistm}), and may indicate electron scattering involving Fermi arc surface states\cite{kwon2020quasiparticle, Yuan_qpi_2018, Zhang_qpi_2017, lin_visualizing_2017}. These features are present in Fig.\,\ref{fig:fig2}e because the tunnelling electrons have energies determined by $V_\mathrm{d.c.}$, which lies within the energy range of the surface states in \ce{WTe2}, including topologically-protected Fermi arc surface states\cite{kwon2020quasiparticle, Yuan_qpi_2018, Zhang_qpi_2017, lin_visualizing_2017}. However, in Fig.\,\ref{fig:fig2}d, most of the lightwave-driven electrons have energies well beyond the surface states near the Fermi level, resulting in faint arc-like features.

\subsection*{Subsection 3: Picometre-scale microscopy}

Since the phase transition is structural in nature\cite{guan_manipulating_2021, Hein2020mode, Soranzio2022, He2016softening, sie_ultrafast_2019, ji_manipulation_2021, Drueke2021, Qi2022intralayer}, $I_\mathrm{Y}$ is a measure of not only the LDOS difference between the two phases, but also the spatial shift of the top atomic plane. This differential measurement mode, illustrated in Fig.\,\ref{fig:fig3}a, allows us to capture details of the phase transition in real space with resolution beyond what is possible with conventional atomically resolved STM\cite{Chen2021}. Subatomic spatial resolution has been achieved with atomic force microscopy\cite{giessibl2000subatomic,welker2012revealing, emmrich2015subatomic}, enabled by the \qty{40}{pm} decay of the interaction potential\cite{welker2012revealing}. Although the tunnel current decays over a longer distance, differential STM images assembled by subtracting the normalized current at moderate tip height from that at lower tip height have also captured picometre-scale features\cite{welker2012revealing}. Critically, our approach to subatomic resolution imaging is intrinsically differential: we measure a changing surface rather than relying on post-processing. By employing lock-in detection, we drastically improve the signal-to-noise ratio of picometre-scale imaging with tunnelling microscopy. Direct differential imaging of the two phases also greatly mitigates effects due to sample drift, which would otherwise obscure subatomic detail in a post-subtraction of the images. 

We demonstrate subatomic spatial resolution of the phase transition via $I_\mathrm{Y}(x,\,y)$ in Fig.\,\ref{fig:fig3}b--d. Figure \ref{fig:fig3}b shows an area encompassing six of the lower tellurium atoms (strong positive signal), while Figs.\,\ref{fig:fig3}c and \ref{fig:fig3}d focus on the subatomic image contrast near the upper tellurium atoms (see Supplementary Fig.\,\ref{sifig:3D_THzSTM} for a comparison of $I_\mathrm{Y}(x,\,y)$ to $I\textsubscript{d.c.}(x,\,y)$). The cross-sections in Fig.\,\ref{fig:fig3}e emphasize the subatomic spatial resolution of $I_\mathrm{Y}$, revealing features with dimensions that are approximately one-tenth of the interatomic spacing. Next, we compare the spatial distribution of $I_\mathrm{Y}$ (Fig.\,\ref{fig:fig3}b) with calculated charge density maps for T\textsubscript{d}-\ce{WTe2} and 1T\textquotesingle{}-\ce{WTe2}. Figure \ref{fig:fig3}f shows separate charge density maps for each of the phases, with each image representing the spatial distribution of the charge density integrated from 0 to \qty{10}{meV} (relative to the Fermi level), following the Tersoff-Hamann approximation\cite{Chen2021} (see Methods). Before subtracting the images, we introduce a relative shift along the $\vec{b}$-axis, which is the direction of shear motion expected during the phase transition. The magnitude of this shift is used as a fit parameter when subtracting the charge densities for each phase, $\rho_{\mathrm{T}_\mathrm{d}}-\rho_\mathrm{1T'}$, arriving at a differential charge density map between the two phases (see Fig.\,\ref{fig:fig3}g and Extended Data Fig.\,\ref{fig:ext-shiftsubtract}). A shift of the top atomic plane by 7~$\pm$~3~pm captures the key features observed in the experimental data (Fig.\,\ref{fig:fig3}b).

Figure \ref{fig:fig3}h shows four layers of the lattice structures for T\textsubscript{d}-\ce{WTe2} and 1T\textquotesingle{}-\ce{WTe2}, with the bottom layer aligned for comparison. Due to the distinct  symmetries of the two phases, the layers shift in pairs across the phase transition, resulting in an approximately commensurate, periodic lattice every 12 layers due to the $\approx$4$^{\circ}$ mismatch between unit cells of T\textsubscript{d} and 1T\textquotesingle{} (see Extended Data Fig.\,\ref{fig:ext-junctionvisualization}). Our calculations find that the relative lateral shift between unpaired layers is $\qty{100}{pm}$ (e.g., between the second and third layers in Fig.\,\ref{fig:fig3}h), whereas it is only a few picometres for paired layers (between the first and second, and between the third and fourth layers in Fig.\,\ref{fig:fig3}h). The close correspondence between our extracted shift of the top atomic layer (Figs.\,\ref{fig:fig3}b,g) and the calculated shifts between paired layers (Extended Data Fig.\,\ref{fig:ext-junctionvisualization}) suggests that our surface (1) is paired with the layer directly beneath it, which remains stationary, and (2) translates in-plane by only 7~$\pm$~3~pm relative to this underlying layer. We emphasize that both interlayer shear motion and intralayer distortion of the surface layer to the 1T\textquotesingle{}-\ce{WTe2} configuration are necessary to reproduce the experimental data, as subtracting shifted charge densities for the T\textsubscript{d} phase alone is insufficient.

\subsection*{Subsection 4: Electronic character of the metastable state}

The electronic band structures of the bulk T\textsubscript{d} and 1T\textquotesingle{} phases calculated by hybrid-level density functional theory (see Methods and Extended Data Fig.\,\ref{fig:ext-banddiagrams}) provide an informative point of comparison for the metastable state observed in our experiments, especially in the plane of the layers ($k_z = 0$). Figures \ref{fig:fig4}a (T\textsubscript{d} phase) and \ref{fig:fig4}b (1T\textquotesingle{} phase) show the bulk band structures in the ($k_x$--$k_y$) plane near the Weyl points (covering only half of the first Brillouin zone due to symmetry) and Fig.\,\ref{fig:fig4}c shows the bulk bands along the $\Gamma$--$k_x$ direction ($\Gamma$--X for T\textsubscript{d} and $\Gamma$--Z for 1T\textquotesingle{}). In the 1T\textquotesingle{}-\ce{WTe2} phase, centrosymmetry enforces band degeneracy, whereas in the T\textsubscript{d}-\ce{WTe2} phase, centrosymmetry breaking lifts this degeneracy (Fig.\,\ref{fig:fig4}c), resulting in band crossings (Weyl points) where the valence and conduction bands touch (Figs.\,\ref{fig:fig4}a,c). Projecting the bulk bands along the surface normal in the $k_z$ direction reveals the range of energies and ($k_x$,\:$k_y$) values that host bulk electronic states, as shown in Extended Data Fig.\,\ref{fig:ext-bulkbandprojections}. The surface electronic bands, including Fermi arc states, occupy the regions of energy and in-plane momentum between the bulk projections\cite{soluyanov2015type}. Thus, a phase transition from T\textsubscript{d}-\ce{WTe2} to 1T\textquotesingle{}-\ce{WTe2}, which restores the bulk band degeneracy, allows surface states to emerge in regions of energy-momentum space that were previously inaccessible. Simultaneously, Fermi arc surface states connecting Weyl points in T\textsubscript{d}-\ce{WTe2} disappear upon annihilation of the Weyl points across the phase transition.

The prominent quasiparticle standing waves observed around surface defects in Fig.\,\ref{fig:fig2}e -- which even change sign between adjacent maxima and minima -- illustrate that $I_\mathrm{Y}$ is dominated by the differences between the surfaces states of the ground and metastable phases. By experimentally tuning $V_\mathrm{d.c.}$, we can therefore explore the surface electronic differences measured by $I_\mathrm{Y}$ for energies near the Fermi level (Fig.\,\ref{fig:fig4}d). We observe a differential signal between the two phases only for biases between \qty{\pm 30}{mV} and find that $I_\mathrm{Y}$ is maximized at energies coinciding with the extrema of the non-degenerate bulk bands that appear only in T\textsubscript{d}-\ce{WTe2} (compare to Fig.\,\ref{fig:fig4}c). Since the bulk bands of each phase define the allowed energies and momenta of the corresponding surface states, this is the energy range where surface bands exhibit the greatest variation between phases and also where Fermi arc surface states reside. The differential current between phases is hence consistent with a topological phase transition of the top atomic layers, where the surface electronic structure shifts from that of T\textsubscript{d}-\ce{WTe2} to one consistent with the surface of 1T\textquotesingle{}-\ce{WTe2}, while the \ce{WTe2} bulk remains in the T\textsubscript{d} phase (see Extended Data Fig.\,\ref{fig:ext-junctionvisualization}). Since topologically-protected Fermi arc surface states necessarily arise from Weyl points in the bulk, we expect these states to shift below the surface in our experiments, relocating to the new interface of T\textsubscript{d}-\ce{WTe2} underneath the shifted layer.


\section*{Discussion}

The ability to locally switch surface topology presents an exciting opportunity for actively controlling electronic transport in quantum materials at the atomic scale. Hybrid-level DFT calculations reveal that the Weyl points in \ce{WTe2} are highly sensitive to subtle changes in the interlayer coupling (see Methods and Extended Data Fig.\,\ref{fig:ext-deformationstrain}), with strain along the $\vec{c}$-axis causing their creation or annihilation, consistent with previous studies\cite{soluyanov2015type, Chang2016, Bruno_observ_2016, Wang_observation_2016, DiSante2017, sie_ultrafast_2019, Rossi2020, Kim2017}. This sensitivity opens avenues for the design of device architectures that can selectively quench topologically protected transport channels, thereby enabling dynamic lightwave control over quantum states. The success of such an approach relies on the spatial confinement and strength of the terahertz near-field at the tip apex, which acts as a localized trigger for inducing phase transitions and governs the lateral extent of the metastable state. Our experiments demonstrate that the terahertz field strength, surface termination and local strain distribution all play crucial roles in the observed phenomena (see Methods for further discussion). 

Critically, THz-STM also enables identification of dynamically modulated surface states in real-space with picometre-scale resolution. Further yet, the prospect of quasiparticle interference imaging with THz-STM promises complementary momentum-space measurements that can help identify the electronic properties of topological materials. Through highly localized lightwave-driven phase transitions, THz-STM uniquely facilitates time-domain exploration of quantum material surface dynamics at the atomic scale. These capabilities establish the groundwork for systems where emergent topological phases can be accessed and controlled\cite{basov2017towards, Torre2021, Ma2021, Bao2022, li2023approaching, Zhai2024}, ultimately enabling novel device architectures that exploit metastable phase transitions to precisely control electron transport and spin currents.

\newpage


\section*{Methods}

\subsection*{Sample preparation}

A single crystal Au(111) sample cleaned by repeated cycles of argon ion bombardment and annealing to \qty{850}{K} was used to acquire the reference waveform for atomic-scale terahertz time-domain spectroscopy\cite{Jelic_atomicTHzTDS_2024} in Fig.\,\ref{fig:fig1}c and prepare the tip apex for all measurements on \ce{WTe2}. The single crystal \ce{WTe2} samples (2Dsemiconductors USA) were grown using the float zone technique to prevent halide contamination, with a confirmed purity of 99.9999\% (6N). The \ce{WTe2} crystals were affixed  to a metallic sample holder plate using conductive epoxy resin. To cleave the sample \textit{in situ}, a ceramic rod was attached to the \ce{WTe2} surface with epoxy and cleaved under ultrahigh-vacuum (UHV) inside the STM chamber at 77 K, revealing an atomically clean surface with a low density of defects and step-edges. A terrace was not directly observed in the STM scan area except for a single instance.

\subsection*{Scanning tunnelling microscope and terahertz optical setup}

In this study, we use a commercial UHV STM system (CreaTec Fischer \& Co. GmbH) with a base pressure of \qty{5e-11}{mbar}. All THz-STM measurements were performed at \qty{8}{K} (STM base temperature with optical access). In the STM scan head, the bias voltage is applied to the sample and the tip is held at virtual ground. A \qty{1}{kHz} bandwidth preamplifier (Femto DLPCA-200) with a gain of $10^9$ V/A is used to detect the tunnel current. The experimental images were acquired with the system in thermal equilibrium, with a lateral drift rate of approximately 1 picometre per minute. Electrochemically etched tungsten tips were prepared in UHV with field-directed argon ion sputtering\cite{schmucker_field_2012} and our tip apex shaping procedure for THz-STM on Au(111), as described in ref.\,[\citen{Jelic_atomicTHzTDS_2024}]. Raw STM data was analysed using Gwyddion\cite{Necas2012}.

Ultrafast terahertz pulses are generated via tilted-pulse front optical rectification\cite{hirori2011single} using a Light Conversion CARBIDE laser system (Yb:KGW) with a repetition rate of \qty{1}{MHz} and pulses at a centre wavelength of \qty{1030}{nm} with a duration of \qty{230}{fs}. The \textit{p}-polarized terahertz pulses pass through three C-cut sapphire windows before entering the STM scanhead and being focused onto the STM tip by a 60$^\circ$ off-axis aluminium parabolic mirror with a focal length of \qty{33.85}{mm} and diameter of \qty{25.4}{mm}. The alignment of the terahertz-pulse-train on the tip apex is preserved when exchanging samples. The optical setup for terahertz pulse generation and detection is further described in refs.\,[\citen{ammerman2021lightwave, Jelic_atomicTHzTDS_2024}].

\subsection*{Lightwave driven scanning tunnelling microscopy and spectroscopy}

To coherently drive tunnelling between the STM tip and sample, lightwave-driven STM operates in the strong-field regime of nonlinear optics\cite{cocker_ultrafast_2013, cocker_nanoscale_2021, Cocker2016, Yoshioka2016, jelic2017ultrafast, Muller2020, Garg2020 , ammerman2021lightwave, Yoshida2021, WangHo2022, Tianwu2024, Siday2024, roelcke2024ultrafast, Bobzien2024, Jelic_atomicTHzTDS_2024, sheng2024terahertz}. Strong electric fields exceeding 1\,V/nm are readily achieved because the tip--sample junction enhances the free space amplitude by a factor of $10^5$ to $10^6$ (eg.\;\qty{100}{V/cm}\,$\times$\,$10^5$ = \qty{1}{V/nm})\cite{Cocker2016, Yoshioka2016, jelic2017ultrafast, peller2021quantitative}. Since terahertz pulses at these fields strengths are well within the quasistatic limit\cite{CockerHegmann2021, jelic2017ultrafast}, the transient electric field acts as a bias voltage that induces an ultrafast tunnel current, enabling terahertz tunnelling spectroscopy\cite{ammerman2021lightwave, sheng2024terahertz, roelcke2024ultrafast, Bobzien2024, Jelic_atomicTHzTDS_2024} and differential terahertz tunnelling spectroscopy\cite{Jelic_atomicTHzTDS_2024}. The voltage calibration (Supplementary Fig.\,\ref{sifig:thzvoltagecalibration}) and temporal profile (Extended Data Fig.\,\ref{fig:ext-validation}) of the terahertz near-field waveform at the tip apex is determined following ref.\,[\citen{Jelic_atomicTHzTDS_2024}]. The calibration constant for our STM tip on \ce{WTe2} is $\alpha$ = (1 V)/(18 V/cm), while on Au(111) it is $\alpha$ = (1 V)/(15 V/cm),  where $\alpha$ = $V$\textsubscript{SF,pk}/$E$\textsubscript{SF,pk} and $F=\alpha/z\textsubscript{0}$ with $z\textsubscript{0}$ as the absolute tip height and $F$ as the field enhancement.

In THz-STM of \ce{WTe2} the field enhancement is critical, as resonantly driving the phase transition requires terahertz pulses with peak amplitudes that are on the order of ten megavolts per centimetre (\qty{1}{V/nm}, while the terahertz field strength incident on the tip is only tens of volts per centimetre. The precise $E\textsubscript{THz,pk}$ required for regime II is determined by both the microscopic tip geometry and the tip-sample distance, as both factors contribute to the overall field enhancement. When operating in regime II, we ensure that the sample has mostly recovered back to the T\textsubscript{d} phase of \ce{WTe2} from the terahertz-field-induced metastable state during the one millisecond window that the terahertz pulse-train is fully blocked by the optical chopper operating at $f\textsubscript{THz}$ = 477 Hz. This recovery is verified by comparing the phase transition behaviour at a reduced chopping frequency (Extended Data Fig.\,\ref{fig:ext-oscope}), ensuring consistent modulation of the field-driven phase transition throughout experiments. The pronounced impact of the terahertz field on the topography observed in regime II is unprecedented in THz-STM, as the peak field is present in the tunnel junction for only a fraction of a picosecond every microsecond, resulting in minimal modulation of the total current (and hence topography).

\subsection*{Atomic-scale terahertz time-domain spectroscopy}

We perform atomic-scale terahertz time-domain spectroscopy (THz-TDS) by dividing each terahertz pulse into a strong-field pulse and a weak-field replica\cite{Jelic_atomicTHzTDS_2024}. The strong-field pulse creates a unipolar current pulse in the junction through lightwave-driven tunnelling. By scanning the temporal delay between the weak-field pulse and the strong-field pulse, a terahertz near-field waveform is recorded for the weak pulse through the time dependence of $I_\mathrm{X}$. The purple (orange) spectrum at the top of Fig.\,\ref{fig:fig1}c shows the spectral amplitude (phase) for the \ce{WTe2} sample. This spectrum was obtained from a Fourier transform of the oscillating near-field in the tunnel junction, divided by a reference spectrum measured on Au(111) to remove the spectral response of antenna coupling to the tip (see Extended Fig.\,\ref{fig:ext-validation} and ref.\,[\citen{Jelic_atomicTHzTDS_2024}] for details). The parameters used during THz-TDS on \ce{WTe2} were $V_\mathrm{d.c.} = \qty{1}{V}$, $I_\mathrm{d.c.} = \qty{100}{pA}$, $E_\mathrm{THz,pk} = \qty{+190}{V/cm}$, $E_\mathrm{WF,pk} = \qty{7}{V/cm}$, while for Au(111) the parameters were $V_\mathrm{d.c.} = \qty{10}{mV}$, $I_\mathrm{d.c.} = \qty{100}{pA}$, $E_\mathrm{THz,pk} = \qty{+150}{V/cm}$, $E_\mathrm{WF,pk} = \qty{7}{V/cm}$, where $E_\mathrm{WF,pk}$ is the peak electric field of the weak-field terahertz pulse used during THz-TDS with THz-STM\cite{Jelic_atomicTHzTDS_2024}. Both THz-TDS and the voltage calibration (Supplementary Fig.\,\ref{sifig:thzvoltagecalibration}) were performed within regime III, as defined in Fig.\,\ref{fig:fig2}.

\subsection*{Density functional theory calculations}

Quantum-mechanical calculations are performed using density functional theory (DFT) as implemented in the CRYSTAL23 package\cite{erba_crystal23_2022}, which uses Gaussian-type basis functions. Our calculations employ the global hybrid PBE0 exchange-correlation functional\cite{adamo_PBE0_1999} and, for comparison, we also carry out calculations with the Perdew-Burke-Ernzerhof (PBE) generalized gradient approximation (GGA)\cite{perdew_PBE_1996} functional. Grimme’s D3 dispersion correction, combined with Becke-Johnson damping, is employed to account for long-range interactions, including van der Waals and London dispersion forces\cite{grimme_D3_2010, grimme_bjdamping_2011}. The direct inversion of the iterative subspace (DIIS) method is used as a convergence accelerator\cite{pulay_diis_1980, pulay_diis_1982}. Spin-orbit coupling (SOC) effects are taken into account within CRYSTAL23 by employing a two-component (Pauli spinor) representation of the wave function\cite{desmarais_2019_spin, desmarais_socfockexchange_2019, desmarais_2020_adiabatic, desmarais_socperturbation_2023, comaskey_soc_2022}. This two-component representation -- also known as spin-current DFT -- is shown to provide accurate, quantitative results for Weyl semimetals\cite{Bodo2022}. The orbitals of tungsten and tellurium are represented using triple-zeta valence with polarization (TZVP) basis sets \texttt{pob\_TZVP\_rev2} for non-SOC calculations\cite{laun_rev2basis_2021} and \texttt{pob\_TZVP\_rev2\_SOC} for SOC calculations\cite{laun_rev2basisrelativistic_2022, desmarais_socCRYSTAL_2020}, which include corrections for basis set superposition error (BSSE). Furthermore, relativistic Stuttgart small-core (STUTSC) effective potentials are employed for tungsten and tellurium during SOC calculations. For geometry optimizations, the dimensionless integral tolerances for the Coulomb and exchange series (ITOL1, ITOL2, ITOL3, ITOL4, ITOL5; see CRYSTAL manual for details) are set to 8, 8, 8, 10, 34, respectively. For single-point energy and SOC calculations, these tolerances are set to 9, 9, 9, 11, 38, respectively.

The lattice constants of the T\textsubscript{d} phase (Pmn2\textsubscript{1}, space group 31) are initialized from the experimental crystallographic data reported by Mar \textit{et al.}\;(ref.\,[\citen{mar_Tdstructure_1992}]), while the lattice constants of the 1T\textquotesingle{} phase (P2\textsubscript{1}/m, space group 11) are initialized from the atomic coordinates and lattice parameters of \ce{MoTe2} (ref.\,[\citen{wang_1TpMoTe2structure_2016}]). A full geometry optimization (atom positions and unit cell parameters) of the T\textsubscript{d} phase is initially carried out using the conventional GGA-PBE functional in order to verify our calculations with previously reported results\cite{soluyanov2015type}. Subsequent calculations employ the global hybrid functional, PBE0, which includes 25\% exact exchange. Full geometry optimizations are then carried out for both the T\textsubscript{d} and 1T\textquotesingle{} phases using the PBE0 functional. The total energy convergence tolerance is set to $2.72\times10^{-7}$ eV to achieve the first level of geometry optimizations. Subsequently, high level single-point energy calculations are carried out with a total energy convergence of $2.72\times10^{-8}$ eV. Additional single point calculations are performed including SOC using these optimized geometries. Reciprocal space is sampled using a 24$\times$24$\times$24 Monkhorst-Pack grid within the reciprocal primitive lattice (24$\times$24$\times$1 for slab calculations). Post-processing of the wavefunctions obtained from these single-point energy calculations enables plotting the band structures and charge densities. Three-dimensional band structure plots are obtained using code by W.\;Comaskey\cite{ComaskeyGithub}. Atomic lattice diagrams are generated using VESTA\cite{Momma2011}.

Two-, three-, and four-layer slab structures are generated by cleaving the optimized bulk crystal structures parallel to the (001) surface. The structures are then repeated within a large vacuum of 500~\AA~to prevent interactions between the slabs. Then, the lattice parameters and atomic positions are fully optimized using the PBE0 functional (without SOC), and band structures were obtained from single-point energy calculations including SOC (see Extended Data Fig.\,\ref{fig:ext-banddiagrams}). The charge density maps are simulated using the Tersoff-Hamann approximation\cite{tersoff1985theory,lee_1998_abinitio} by projecting surface charge densities within an energy range of 0 to 10~meV at a distance of 2\AA~away from the slab. This reduced distance compensates for the DFT simulations’ tendency to underestimate the charge distribution into the vacuum due to localized basis states\cite{ammerman2021lightwave}. Energy-projected charge densities with SOC are not implemented in the current version of the CRYSTAL software; therefore, non-SOC charge densities are shown in this work. The surface charge densities in Fig.\,\ref{fig:fig3}f are calculated from the three-layer slabs. 

For all previously mentioned calculations, the convergence threshold for root mean squared (RMS) forces, maximum forces, RMS atomic displacements, and maximum atomic displacements is set to $1.54\times10^{-2}$ eV \AA$^{-1}$, $2.31\times10^{-2}$ eV \AA$^{-1}$, $6.35\times10^{-4}$ \AA, and $9.53\times10^{-4}$ \AA, respectively. On the other hand, the phonon frequencies shown in Table~\ref{tab:sitab-phonondispersion} are calculated using a supercell approach\cite{pascale_2004_calculation, zicovich_2004_calculation} that breaks translational symmetry. Vibrational frequencies are highly sensitive to structural parameters, so stricter convergence criteria are set for optimizations prior to these calculations. The stricter convergence threshold for RMS forces, maximum forces, RMS atomic displacements, and maximum atomic displacements is set to $1.54\times10^{-3}$ eV \AA$^{-1}$, $2.31\times10^{-3}$ eV \AA$^{-1}$, $6.35\times10^{-5}$ \AA, and $9.53\times10^{-5}$ \AA, respectively. The energy threshold is also stricter, at $2.72\times10^{-11}$ eV. Phonon dispersion is calculated for a two-layer T\textsubscript{d}-\ce{WTe2} slab (Extended Data Fig.\,\ref{fig:ext-phononbandstructure}), using a $2\times2$ supercell. At the $\Gamma$ point, the vibrational mode at 0.29 THz corresponds to an interlayer shear mode, indicated by the atomic movement predominantly in the $\vec{b}$-direction related with this eigenvector.

\subsection*{Extracting the interplanar shear amplitude}

The terahertz-pulse-induced phase transition in \ce{WTe2} is strongly influenced by the broken inversion symmetry of the T\textsubscript{d} phase, which results in two distinct surface terminations\cite{Bruno_observ_2016, Rossi2020, wu_observ_2016, SanchezBarriga2016, Yuan_qpi_2018, Wang_observation_2016}, labelled \textquotesingle{}Surface A\textquotesingle{} and \textquotesingle{}Surface B\textquotesingle{}. The strong terahertz field driving the phase transition is concentrated at the surface layer, decaying rapidly inside the sample and radially away from the tip\cite{peller2021quantitative}. In our experiments, nearly the entire \ce{WTe2} crystal remains in the T\textsubscript{d} phase, except for a nanoscale area near the tip apex where terahertz pulses induce a metastable state on Surface A, shifting the surface layer by 7~$\pm$~3~pm into a structure resembling 1T\textquotesingle{}-\ce{WTe2} (Extended Data Fig.\,\ref{fig:ext-shiftsubtract}). In T\textsubscript{d}-\ce{WTe2}, Surface A is structurally paired with its adjacent subsurface layer, exhibiting a relative lateral shift of only a few picometres between corresponding lattice sites in the T\textsubscript{d} versus 1T\textquotesingle{} phases (Extended Data Fig.\,\ref{fig:ext-junctionvisualization}a). Our extracted experimental shift for the surface layer relative to the tip apex (7~$\pm$~3~pm) confirms that the second layer remains stationary; otherwise, the surface layer would shift by approximately $\qty{100}{pm}$, which we do not observe experimentally. 

In contrast, for Surface B the interfacial layer is unpaired, which means that driving the surface layer of T\textsubscript{d}-\ce{WTe2} into a metastable phase resembling 1T\textquotesingle{}-\ce{WTe2} requires a \qty{100}{pm} shift relative to the subsurface layer (Extended Data Fig.\,\ref{fig:ext-junctionvisualization}b). During our THz-STM measurements, the terahertz-pulse-induced phase transition is not consistently observed across all regions of the sample, often requiring multiple attempts to find an area that exhibits the effect. This inconsistency may result from either the need for a particular structural strain, or occasionally landing on Surface B, where attempts to locally drive a 100-pm lateral shift between adjacent layers are strongly suppressed by the surrounding atomic lattice, which remains structurally unperturbed due to insufficient local terahertz field strength. Notably, a large interplanar shift has been realized in few-layer \ce{WTe2} by applying static fields on the order of volts per nanometer through large-contact-area electrodes, triggering a polarity switch in the out-of-plane ferroelectric dipole\cite{fei2018ferroelectric, Yang2018, Ni2020, Xiao2020}.

Calculations using the PBE functional\cite{Rossi2020} have suggested that Surface B has Fermi arc surface states residing below the Fermi level, making them accessible for ARPES studies, whereas on Surface A, these states straddle the Fermi level. However, the precise energies and momenta of Weyl points in \ce{WTe2} (and consequently the Fermi arc surface states) vary with external factors such as strain or doping. Additionally, the choice of functional affects the presence and positions of Weyl points, as shown in our calculations. Previous DFT calculations predominantly used the PBE functional and often did not relax the unit cell or atomic coordinates (see Supplementary Tables \ref{sitab:Td_refs}, \ref{sitab:Td_refs2} and \ref{sitab:1Tp_refs}). Our DFT results emphasize the importance of: (1) optimizing lattice structures and (2) using functionals that incorporate a portion of exact exchange when studying Weyl semimetals\cite{Bodo2022} and emergent metastable phases.

\newpage

\section*{Acknowledgements}

We thank W. W. Pai, K. Cleland-Host, S. E. Ammerman, T. Hickle and A. Hayes for scientific support and discussions. Further, we thank R. Loloee and R. Bennett for technical support. This project was supported financially by the Army Research Office (grant no. W911NF2110153) and the Air Force Office of Scientific Research (FA2386-24-1-4042). D.M.-L. acknowledges funding by the National Science Foundation Graduate Research Fellowship under Grant No. 2235783. J.L.M.-C. thanks start-up funds from Michigan State University. This work was supported in part through computational resources and services provided by the Institute for Cyber-Enabled Research at Michigan State University.

\section*{Author contributions}

Experiments and data analysis were carried out by V.J.\,and S.A.\,with support from M.H.\,and T.L.C. Samples and tips were prepared by V.J.\,and M.H. Theoretical quantum calculations were performed by D.M.-L.\,and I.A.B., with support from J.L.M.-C. Interpretation of the results and preparation of the manuscript were carried out by V.J., S.A., D.M.-L., J.L.M.-C.\,and T.L.C. The project was supervised and directed by J.L.M.-C.\,and T.L.C., and the study was conceived by T.L.C.

\section*{Data availability}

The raw data that support the findings of this study are available from the corresponding authors upon reasonable request.

\section*{Code availability}

The code that supports the plots and data analysis of this study are available from the corresponding authors upon reasonable request.

\section*{Competing interests}

The authors declare no competing interests.

\newpage

\newpage

\begin{figure}
    \centering
    \includegraphics[width=1\linewidth]{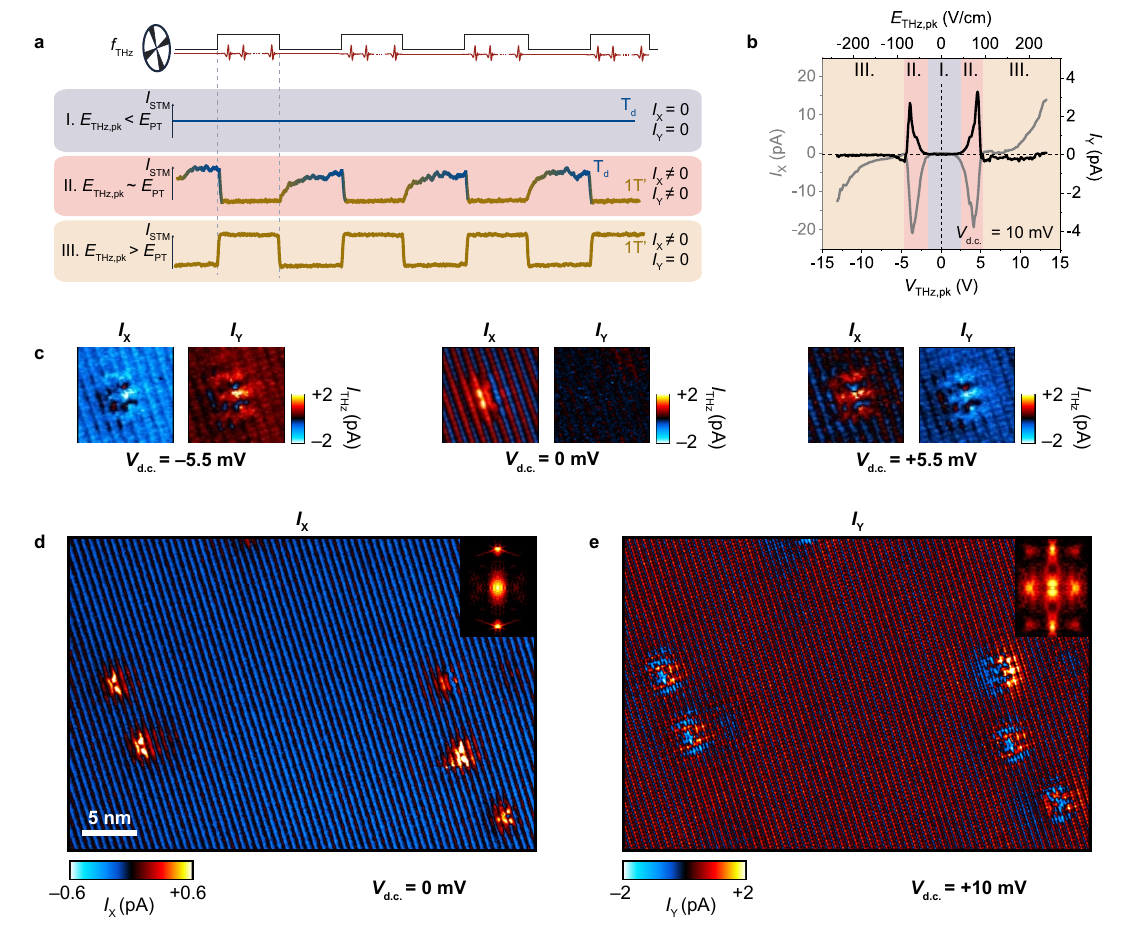}
    \caption{
    \textbf{Figure 2 $\mid$ Differential atomic imaging of a terahertz-field-driven phase transition.} 
    \textbf{a}, Schematic showing the different interaction regimes of \ce{WTe2} within a THz-STM tunnel junction while an optical chopper modulates the incident terahertz-pulse-train (top). Experimental oscilloscope traces are shown for the three interaction regimes. Regime~I: $E_\mathrm{THz,pk}$ does not drive any measurable effects as the tip-enhanced terahertz electric field strength lies below the phase transition onset ($E_\mathrm{PT}$). Regime~II: $E_\mathrm{THz,pk}$ minimally drives tunnelling but is sufficient to induce a phase transition between T\textsubscript{d}-\ce{WTe2} and 1T\textquotesingle{}-\ce{WTe2}, which can be detected via the \ang{90} out-of-phase component of the modulated tunnel current ($I_\mathrm{Y} \neq 0$) when a small d.c.\;bias is also present. Regime~III: terahertz-driven tunnelling becomes dominant, while the differential current, $I_\mathrm{Y}$, between T\textsubscript{d}-\ce{WTe2} and 1T\textquotesingle{}-\ce{WTe2} subsides, since a metastable state emerges as the primary phase at the surface near the tip apex.
    \textbf{b}, $I_\mathrm{THz}$-$V_\mathrm{THz,pk}$ curve on the \ce{WTe2} surface acquired at $V_\mathrm{d.c.} = \qty{-10}{mV}$ and $z = z_0$ with the tip height, $z_0$, set by $V_0 = \qty{10}{mV}$, $I_0 = \qty{100}{pA}$. The solid grey line shows the in-phase component ($I_\mathrm{X}$), while the solid black line shows the out-of-phase component ($I_\mathrm{Y}$). 
    \textbf{c}, THz-STM images of a $\qty{5}{nm} \times \qty{5}{nm}$ area showing both $I_\mathrm{X}(x,y)$ and $I_\mathrm{Y}(x,y)$ for a \ce{WTe2} surface defect acquired at $V_\mathrm{d.c.} = \qty{-5.5}{mV}$ (left), $V_\mathrm{d.c.} = \qty{0}{V}$ (middle) and $V_\mathrm{d.c.} = \qty{+5.5}{mV}$ (right). The left and right image pairs were acquired at $I_\mathrm{d.c.} = \qty{100}{pA}$ and $V_\mathrm{THz,pk} = \qty{-3.3}{V}$, while the middle image pair was acquired at $V_\mathrm{THz,pk} = \qty{-2.8}{V}$ and the tip was approached 150~pm to enhance lightwave-driven tunnelling ($z = z_0 - \qty{150}{pm}$).
    \textbf{d}, Conventional THz-STM rectified current map, $I_\mathrm{X}(x,y)$, of a $\qty{42}{nm} \times \qty{28}{nm}$ area acquired at constant height with $V_\mathrm{d.c.} = \qty{0}{V}$, $V_\mathrm{THz,pk} = \qty{-1.7}{V}$ and $z = z_0 - \qty{50}{pm}$. The tip height $z_0$ was set by $V_\mathrm{d.c.} = \qty{10}{mV}$, $I_\mathrm{d.c.} = \qty{100}{pA}$.
    \textbf{e},~Out-of-phase THz-STM image, $I_\mathrm{Y}(x,y)$, of the same area acquired at $V_\mathrm{d.c.} = \qty{10}{mV}$, $V_\mathrm{THz,pk} = \qty{-2.2}{V}$ and $I_\mathrm{d.c.} = \qty{100}{pA}$. Insets: Symmetrized 2D Fourier transforms with dimensions $\qty{3}{nm^{-1}} \times \qty{4}{nm^{-1}}$ (non-symmetrized are shown in Supplementary Fig.\,\ref{sifig:THzSTMqpi}). Scan speed $\qty{4.3}{nm/s}$ for \textbf{c}-\textbf{e}. A \qty{100}{pm} full-width-half-maximum (FWHM) 2D Gaussian was convolved with the images in \textbf{c}-\textbf{e} for noise reduction.  
    }
    \label{fig:fig2}
\end{figure}

\clearpage

\begin{figure}
    \centering
    \includegraphics[width=1\linewidth]{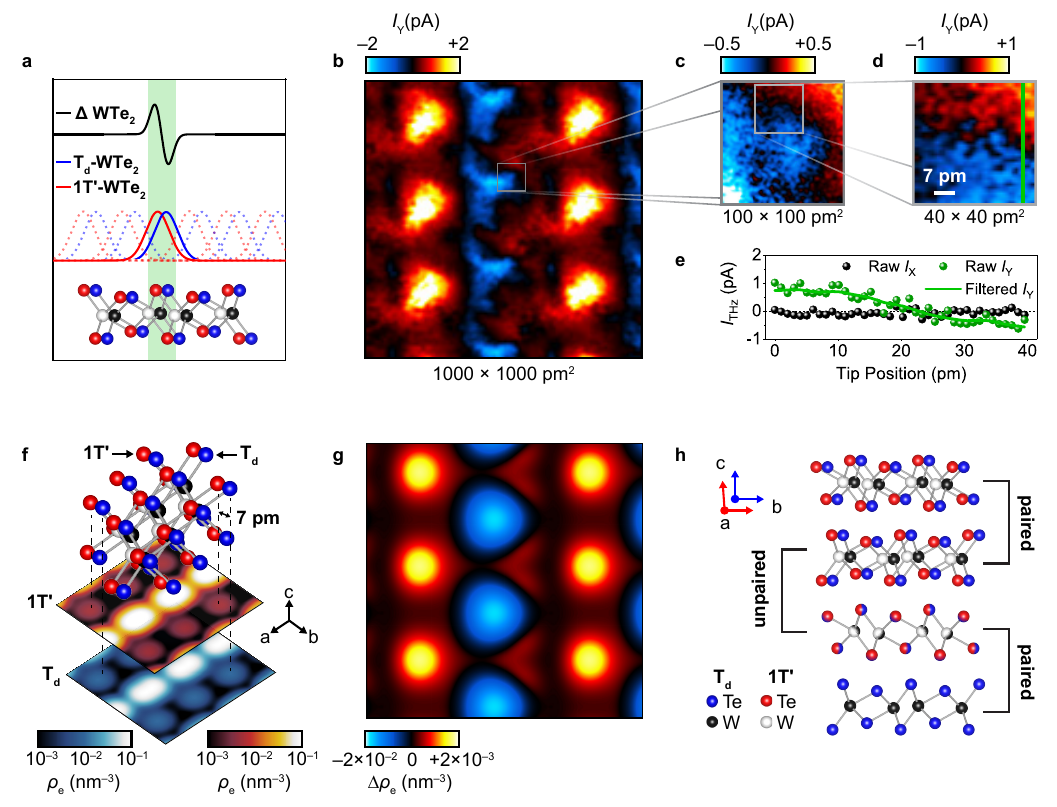}
    \caption{
    \textbf{Figure 3 $\mid$ Picometre-scale microscopy of the terahertz-driven phase transition.} 
    \textbf{a}, Schematic showing how picometre-scale differences in the local electronic properties between T\textsubscript{d} (blue line) and 1T\textquotesingle{} (red line) phases in \ce{WTe2} lead to subatomic resolution microscopy via $I_\mathrm{Y}$ (black line). 
    \textbf{b}-\textbf{d}, Constant-height THz-STM images, $I_\mathrm{Y}(x,y)$, acquired at $V_\mathrm{d.c.} = \qty{10}{mV}$, $V_\mathrm{THz,pk} = \qty{-1.4}{V}$ and tip heights of $z = z_0 - \qty{30}{pm}$ (\textbf{b}, \textbf{c}) and $z = z_0 - \qty{35}{pm}$ (\textbf{d}), with $z_0$ set by $V_\mathrm{d.c.} = \qty{10}{mV}$ and $I_\mathrm{d.c.} = \qty{100}{pA}$. Scan speed \qty{150}{pm \per s} (\textbf{b}), \qty{20}{pm/s} (\textbf{c}) and \qty{10}{pm/s} (\textbf{d}); sampling density \qty{15}{pm/pixel} (\textbf{b}) and \qty{1}{pm/pixel} (\textbf{c},\textbf{d}); Gaussian smoothing filter full-width-half-maximum \qty{15}{pm} (\textbf{b}), \qty{3}{pm} (\textbf{c}), \qty{2}{pm} (\textbf{d}). 
    \textbf{e}, Vertical cross-section of $I_\mathrm{X}$ (black circles) and $I_\mathrm{Y}$ (green circles) indicated by the solid green line in \textbf{d}. 
    \textbf{f}, Simulated charge densities for T\textsubscript{d} (bottom) and 1T\textquotesingle{} (middle) with a schematic (top) showing the atomic positions for each phase (the atomic lattice separation of \qty{7}{pm} is amplified for visual clarity).
    \textbf{g}, Difference image between the simulated charge densities of T\textsubscript{d} and 1T\textquotesingle{} for a $\qty{1000}{pm} \times \qty{1000}{pm}$ square area ($\rho_{\mathrm{T}_\mathrm{d}}$ --$\rho_\mathrm{1T'}$).
    \textbf{h}, Four layers of the orthorhombic T\textsubscript{d} phase (blue and black spheres) and monoclinic 1T\textquotesingle{} phase (red and white spheres). When the lattices of both phases are overlaid, the layers can be grouped with regard to their relative shift between the phases along the $\vec{b}$-axis, as illustrated. The atoms in the bottom layer of both phases are aligned to the same positions within 0.3 pm (only the T\textsubscript{d} phase atoms are visible).
    } 
    \label{fig:fig3}
\end{figure}

\clearpage

\begin{figure}
    \centering
    \includegraphics[width=0.65\linewidth]{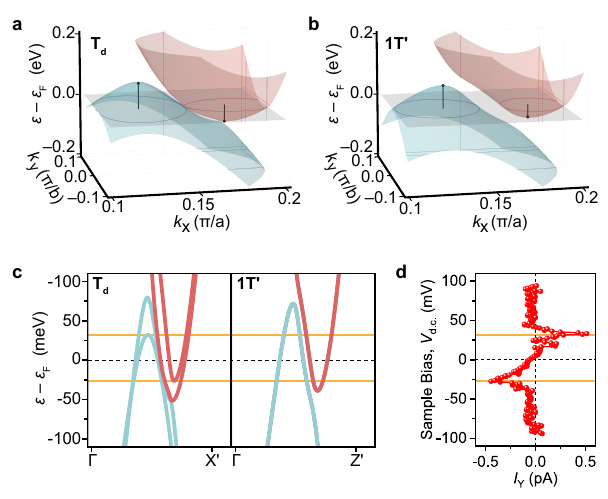}
    \caption{
    \textbf{Figure 4 $\mid$ Identifying a topological symmetry switch in \ce{WTe2}.} 
    \textbf{a}, \textbf{b}, Three-dimensional visualization of the band structure for optimized bulk T\textsubscript{d} (\textbf{a}) and 1T\textquotesingle{} (\textbf{b}) phases along the $k_x-k_y$ plane, near the band crossing. 
    \textbf{c}, Band structure along the high symmetry direction, $\Gamma - \mathrm{X'}$, for the T\textsubscript{d} phase and the equivalent path, $\Gamma - \mathrm{Z'}$, for the 1T\textquotesingle{} phase (see Extended Data Fig.\,\ref{fig:ext-banddiagrams} for the high-symmetry paths). 
    \textbf{d}, Out-of-phase component of the measured tunnel current ($I_\mathrm{Y}$) as a function of $V_\mathrm{d.c.}$, covering the range where topologically protected Fermi arc surface states reside in bulk T\textsubscript{d}-\ce{WTe2}. The terahertz peak voltage is set to $V_\mathrm{THz,pk} = 4.0$ V, ensuring that the field strength underneath the tip apex and in the surrounding area induces a phase transition from T\textsubscript{d}-\ce{WTe2} to the metastable state resembling 1T\textquotesingle{}-\ce{WTe2}. The feedback loop was opened at $V\textsubscript{d.c.}$~=~10~mV and $I\textsubscript{d.c.}$~=~100~pA. The gold horizontal lines in \textbf{c} and \textbf{d} mark the energy positions of the vertices (local extrema) of the non-degenerate valence and conduction bands that are present only in the T\textsubscript{d} phase. The two pairs of Weyl points occur at ($\epsilon$,~$k\textsubscript{x}$,~$k\textsubscript{y}$)~=~($-$12.6,~$\pm$\,0.1516,~0) and ($\epsilon$,~$k\textsubscript{x}$,~$k\textsubscript{y}$)~=~($-$51.1,~$\pm$\,0.1707,~0) with $\epsilon$, $k\textsubscript{x}$ and $k\textsubscript{y}$ in units of meV, $\pi/a$ and $\pi/b$, respectively. 
    }
    \label{fig:fig4}
\end{figure}

\clearpage

\section*{Extended Data}

\begin{extendeddatafigure}
    \centering
    \includegraphics[width=0.9\linewidth]{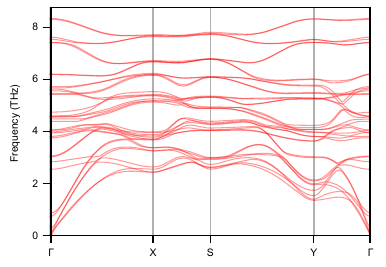}
    \caption{
    \textbf{Extended Data Figure 1 $\mid$ Phonon dispersion of T\textsubscript{d}-\ce{WTe2} calculated using DFT with the hybrid PBE0 functional.}
    Calculated phonon band structure of a two-layer slab along the $\Gamma-$X$-$S$-$Y$-\Gamma$ path. See Extended Data Fig.\,\ref{fig:ext-banddiagrams}a for a schematic of the Brillouin zone.
    }
    \label{fig:ext-phononbandstructure}
\end{extendeddatafigure}

\clearpage

\begin{extendeddatafigure}
    \centering
    \includegraphics[width=1\linewidth]{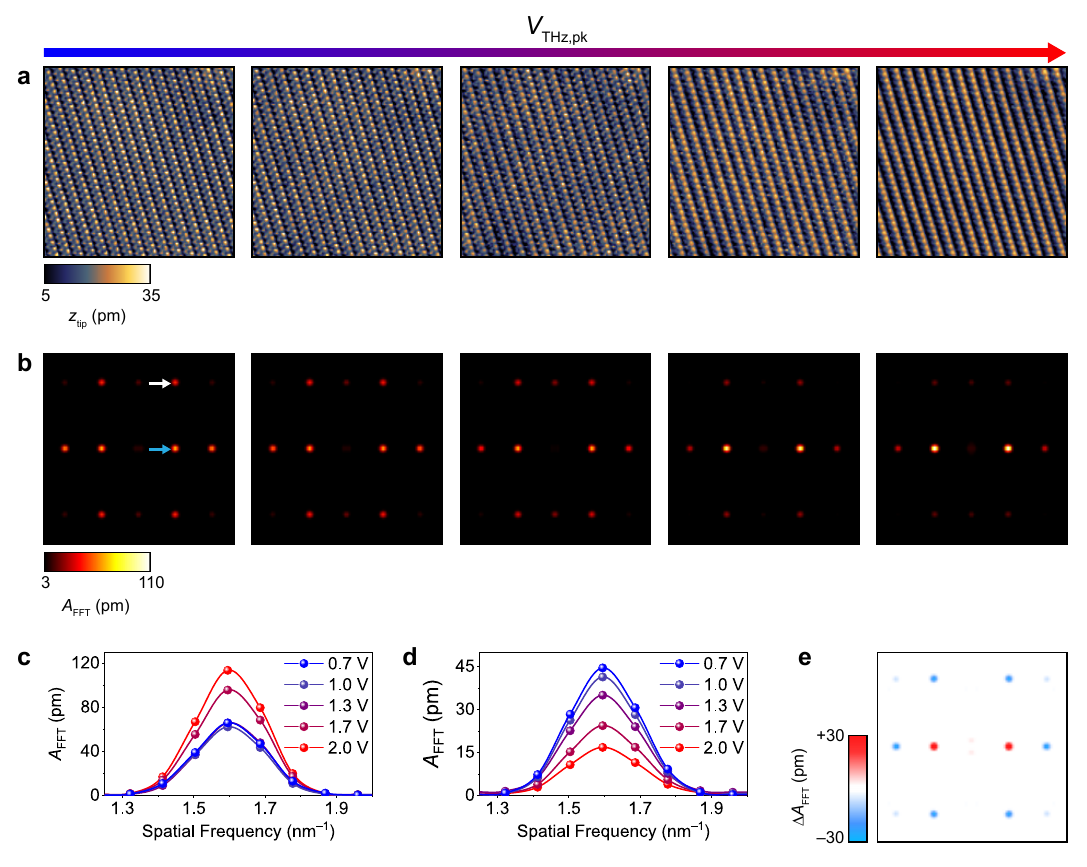}
    \caption{
    \textbf{Extended Data Figure 2 $\mid$ Fourier analysis of \ce{WTe2} STM topography under terahertz pulse illumination.}
    \textbf{a}, STM topography acquired at $V_\mathrm{d.c.} = \qty{10}{mV}$ and $I_\mathrm{d.c.} = \qty{100}{pA}$ with $V_\mathrm{THz,pk}$ at \qty{0.7}{V} (left), \qty{1.0}{V} (middle left), \qty{1.3}{V} (middle), \qty{1.7}{V} (middle right) and \qty{2.0}{V} (right). Chopper frequency 477 Hz; image size $\qty{10}{nm} \times \qty{10}{nm}$; scan speed \qty{4.3}{nm/s}. 
    \textbf{b}-\textbf{d}, Symmetrized 2D Fourier transforms ($\qty{8}{nm^{-1}} \times \qty{8}{nm^{-1}}$) of the images in \textbf{a} with light blue (\textbf{c}) and white (\textbf{d}) arrows denoting Fourier peaks where horizontal cross-sections are shown in \textbf{c} (light blue arrow) and \textbf{d} (white arrow). 
    \textbf{e}, Difference image between the right and left 2D Fourier transforms in \textbf{b}. 
    }
    \label{fig:ext-fourieranalysis}
\end{extendeddatafigure}

\clearpage

\begin{extendeddatafigure}
    \centering
    \includegraphics[width=1\linewidth]{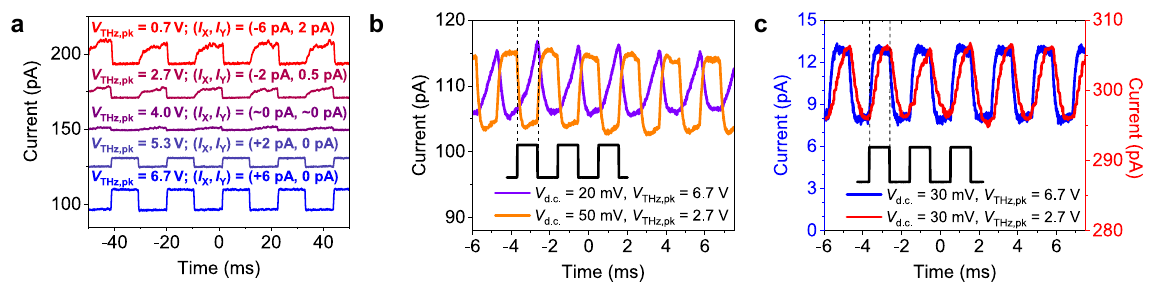}
    \caption{
    \textbf{Extended Data Figure 3 $\mid$ Oscilloscope traces of the total tunnel current.} 
    \textbf{a}, A set of time traces acquired at $f_\mathrm{THz} = \qty{45}{Hz}$, $V_\mathrm{d.c.} = \qty{30}{mV}$ and $I_\mathrm{d.c.} = \qty{100}{pA}$. The traces are vertically offset for clarity. 
    \textbf{b}, A pair of time traces acquired at $V_\mathrm{d.c.} = \qty{20}{mV}$, $I_\mathrm{d.c.} = \qty{110}{pA}$, $V_\mathrm{THz,pk} = \qty{6.7}{V}$ (purple curve) and $V_\mathrm{d.c.} = \qty{50}{mV}$, $I_\mathrm{d.c.} = \qty{110}{pA}$, $V_\mathrm{THz,pk} = \qty{2.7}{V}$ (orange curve), and a square-wave modulation of the terahertz-pulse-train at $f_\mathrm{THz} = \qty{477}{Hz}$ (black curve). Orange curve ($I\textsubscript{X}$,\,$I\textsubscript{Y}$) = (+5~pA,\,0~pA); purple curve ($I\textsubscript{X}$,\,$I\textsubscript{Y}$) = ($-2$~pA,\,$+2$~pA).
    \textbf{c}, A pair of time traces acquired at $V_\mathrm{d.c.} = \qty{30}{mV}$, $I_\mathrm{d.c.} = \qty{10}{pA}$, $V_\mathrm{THz,pk} = \qty{6.7}{V}$ (blue curve) and $V_\mathrm{d.c.} = \qty{30}{mV}$, $I_\mathrm{d.c.} = \qty{300}{pA}$, $V_\mathrm{THz,pk} = \qty{2.7}{V}$ (red curve), and a square-wave modulation of the terahertz-pulse-train at $f_\mathrm{THz} = \qty{477}{Hz}$ (black curve). Red curve ($I\textsubscript{X}$,\,$I\textsubscript{Y}$) = ($+2$~pA,\,$-1.5$~pA); blue curve ($I\textsubscript{X}$,\,$I\textsubscript{Y}$) = ($+2$~pA,\,$0$~pA). 
    }
    \label{fig:ext-oscope}
\end{extendeddatafigure}

\clearpage

\begin{extendeddatafigure}
    \centering
    \includegraphics[width=1\linewidth]{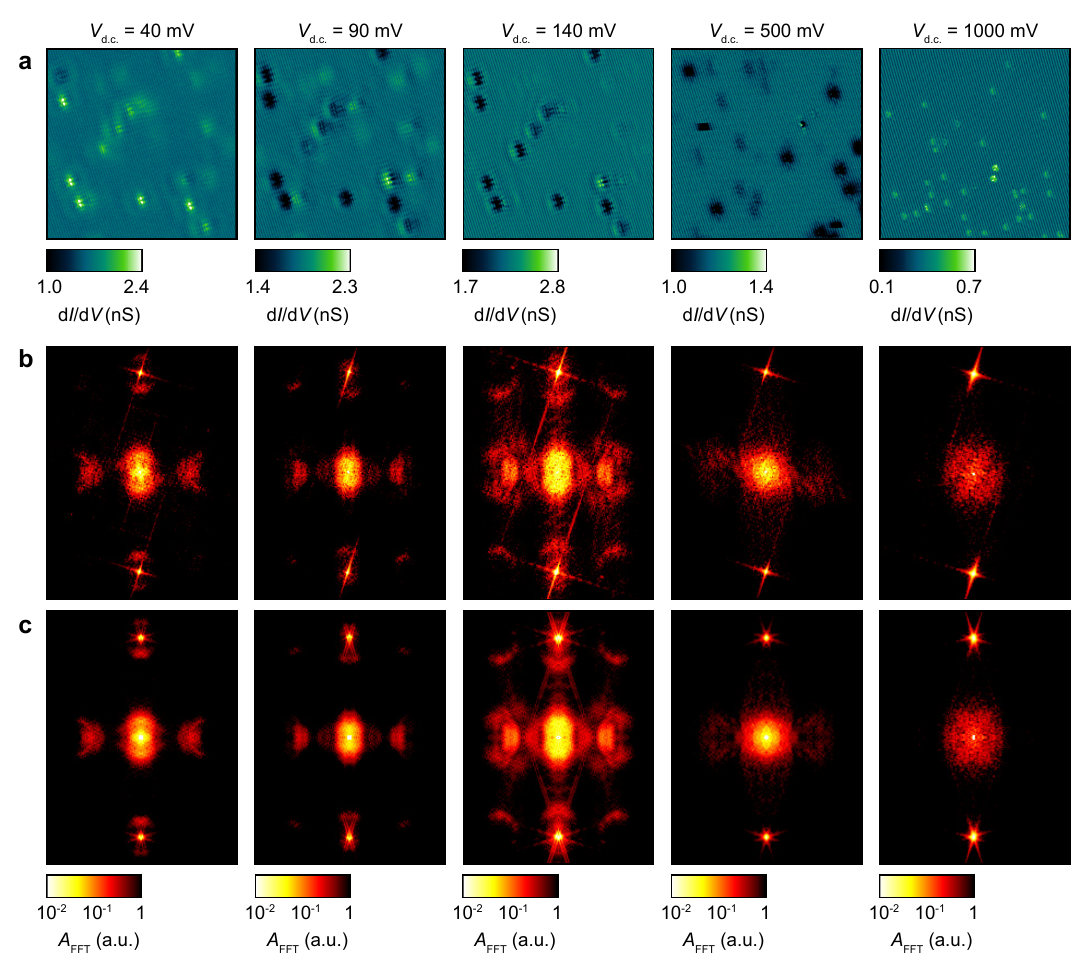}
    \caption{
    \textbf{Extended Data Figure 4 $\mid$ Atomic-scale conductance mapping and FT-STM of \ce{WTe2}.} 
    \textbf{a}, Spatially-resolved conductance (d\textit{I}/d\textit{V}) maps of a \ce{WTe2} surface acquired at a constant-current of $I_\mathrm{d.c.} = \qty{100}{pA}$ and $V_\mathrm{d.c.}$ set to \qty{40}{mV} (left), \qty{90}{mV} (middle left), \qty{140}{mV} (middle), \qty{500}{mV} (middle right) and \qty{1000}{mV} (right). Image size $\qty{50}{nm} \times \qty{50}{nm}$; scan speed \qty{4.4}{nm/s}; $V_\mathrm{a.c.} = \qty{5}{mV}$ (left, middle left, middle) and $V_\mathrm{a.c.} = \qty{10}{mV}$ (middle right, right). 
    \textbf{b}, Fourier-transform STM (FT-STM) image for each respective conductance map positioned above in \textbf{a}. Image size $\qty{3}{nm^{-1}} \times \qty{4}{nm^{-1}}$. 
    \textbf{c}, Symmetrized FT-STM images for each respective image positioned above in \textbf{b}. The cross-hatch features surrounding bright peaks in \textbf{b} and \textbf{c} are inevitable artifacts of the windowless 2D Fourier transforms that were performed on the images in \textbf{a}.
    }
    \label{fig:ext-qpistm}
\end{extendeddatafigure}

\clearpage

\begin{extendeddatafigure}
    \centering
    \includegraphics[width=1\linewidth]{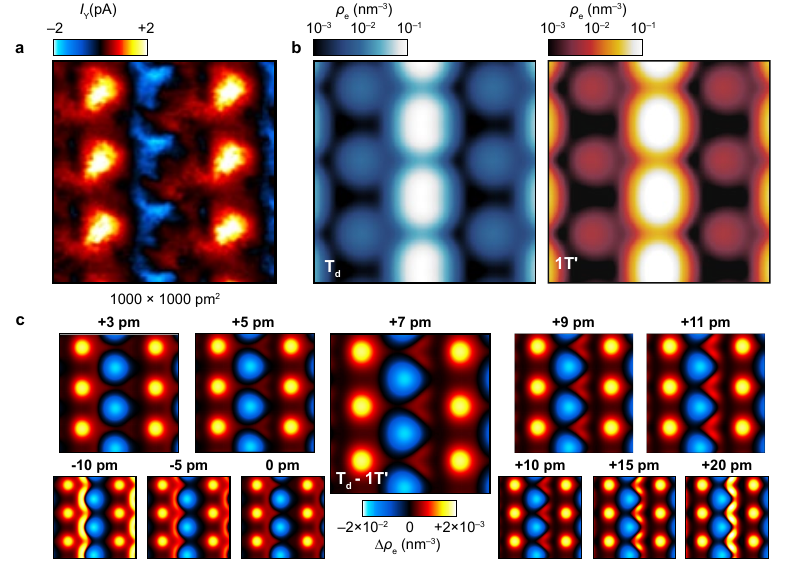}
    \caption{
    \textbf{Extended Data Figure 5 $\mid$ Extracting the picometre-scale shift of the \ce{WTe2} surface layer.}
    \textbf{a}, Experimentally acquired $I\textsubscript{Y}(x,y)$ map from Fig\,\ref{fig:fig3}b.
    \textbf{b}, Surface charge density integrated from 0 to 10 meV above the Fermi level calculated using hybrid-DFT for the T\textsubscript{d} (left) and 1T\textquotesingle{} (right) phases of \ce{WTe2}. 
    \textbf{c}, Subtracted charge densities ($\rho_{\mathrm{T}_\mathrm{d}}-\rho_\mathrm{1T'}$ from \textbf{b}) are shown with the horizontal offset indicated above each image. A shift of \qty{7}{pm} agrees best with the data in \textbf{a} and is shown in the centre. We estimate a tolerance of $\pm \qty{3}{pm}$ based on the reasonable agreement of the charge density differences for an offset of \qty{+4}{pm} to \qty{+10}{pm} (see upper row). The bottom row displays larger shifts in both directions as well as the subtraction without an offset (0 pm), demonstrating that the shift is confined to $7\pm 3\, \mathrm{pm}$ to match the measurement in \textbf{a}.
    }
    \label{fig:ext-shiftsubtract}
\end{extendeddatafigure}

\clearpage

\begin{extendeddatafigure}
    \centering
    \includegraphics[width=1\linewidth]{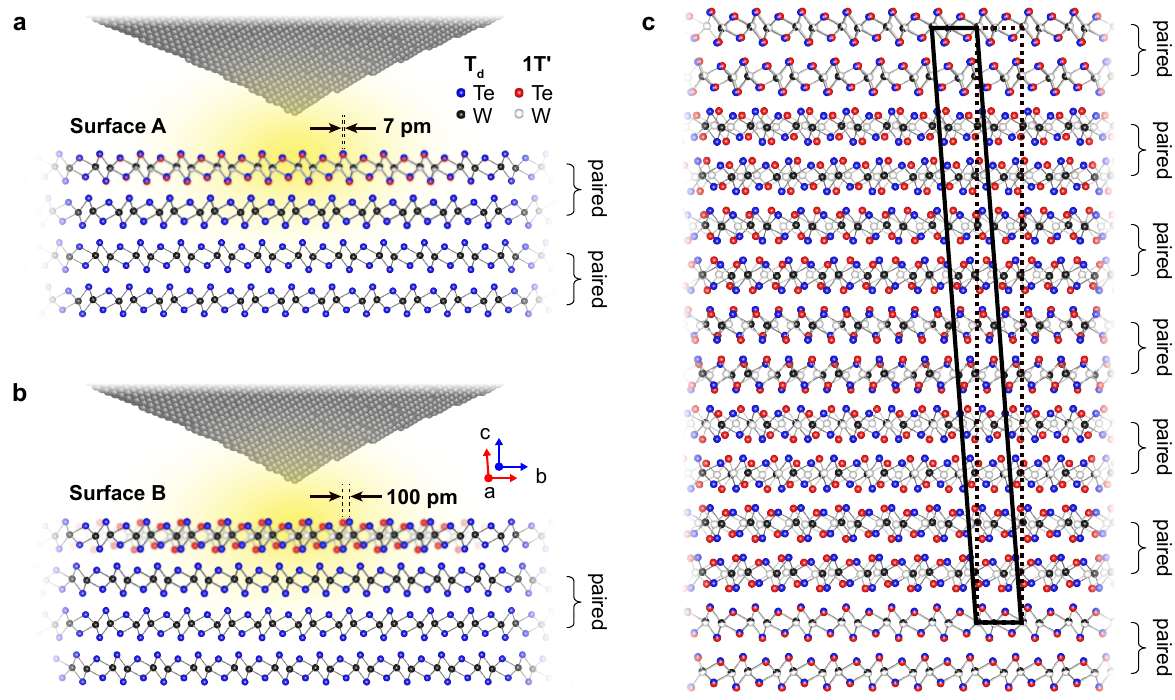}
    \caption{
    \textbf{Extended Data Figure 6 $\mid$ Relaxed atomic coordinates from DFT calculations for both phases of \ce{WTe2}.} 
    Side views are shown along the $\vec{a}$-axis of T\textsubscript{d}-\ce{WTe2} (blue and black spheres) and 1T\textquotesingle{}-\ce{WTe2} (red and white spheres). 
    \textbf{a}, On Surface A, the top layer shifts by approximately \qty{7}{pm} relative to the adjacent paired layer beneath it. 
    \textbf{b}, On Surface B, the top layer shifts by approximately \qty{100}{pm} relative to the adjacent unpaired layer beneath it.
    \textbf{c}, The T\textsubscript{d} and 1T\textquotesingle{} lattices align approximately commensurate every 12 layers. The solid black parallelogram highlights the 4$^{\circ}$ tilt of the supercell for 1T\textquotesingle{} with respect to T\textsubscript{d} (dashed black rectangle).
    }
    \label{fig:ext-junctionvisualization}
\end{extendeddatafigure}

\clearpage

\begin{extendeddatafigure}
    \centering
    \includegraphics[width=\linewidth]{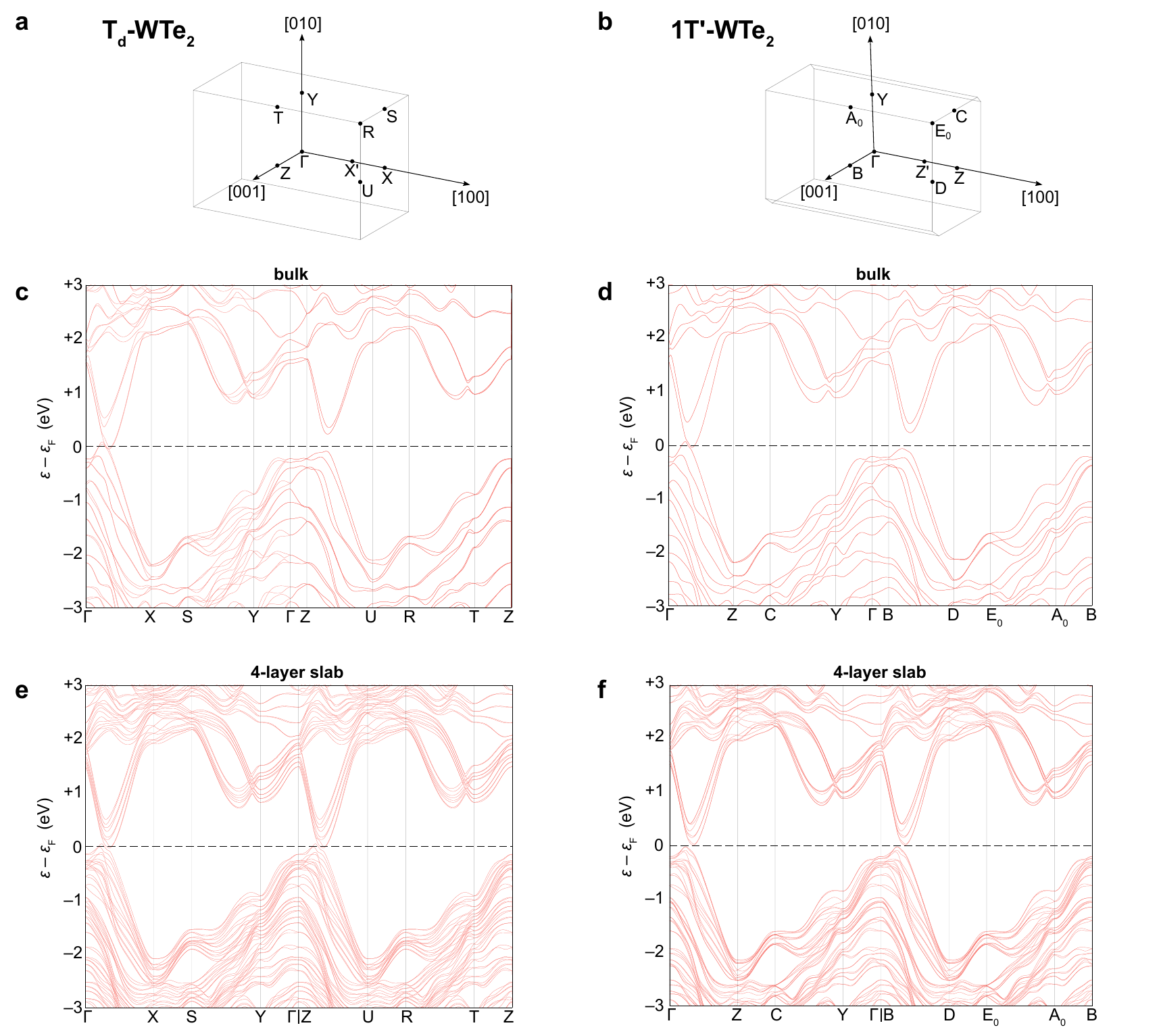}
    \caption{
    \textbf{Extended Data Figure 7 $\mid$ Calculated band structure for \ce{WTe2} using DFT with the PBE0 hybrid functional.}
    \textbf{a},\textbf{b},~Schematic bulk Brillouin zone for orthorhombic T\textsubscript{d}-\ce{WTe2} (\textbf{a}) and monoclinic 1T\textquotesingle{}-\ce{WTe2} (\textbf{b}) showing the high symmetry points. 
    \textbf{c}-\textbf{f}, Band structure for bulk (\textbf{c},\textbf{d}) and four-layer-slab (\textbf{e},\textbf{f}) calculations for T\textsubscript{d}-\ce{WTe2} (\textbf{c},\textbf{e}) and 1T\textquotesingle{}-\ce{WTe2} (\textbf{d},\textbf{f}).
    }
    \label{fig:ext-banddiagrams}
\end{extendeddatafigure}

\clearpage

\begin{extendeddatafigure}
    \centering
    \includegraphics[width=0.65\linewidth]{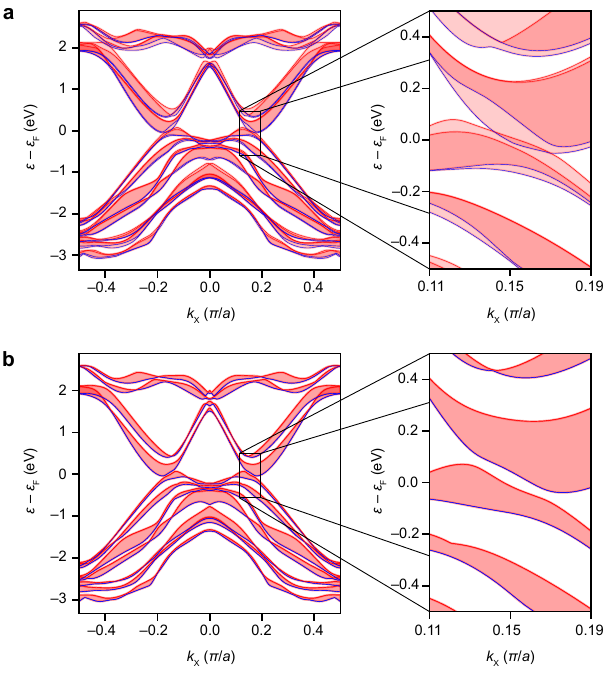}
    \caption{
    \textbf{Extended Data Figure 8 $\mid$ Projection of the bulk electronic states of \ce{WTe2} along the $\vec{c}$-axis.} \textbf{a},\textbf{b},~Calculations using hybrid-level DFT of T\textsubscript{d}-\ce{WTe2} (\textbf{a}) and 1T\textquotesingle{}-\ce{WTe2} (\textbf{b}) bulk states shown along the $k\textsubscript{x}$ direction ($\Gamma$--X for T\textsubscript{d} and $\Gamma$--Z for 1T\textquotesingle{}) and projected along the surface normal, i.e., the $k\textsubscript{z}$ direction ($\Gamma$--Z for T\textsubscript{d} and $\Gamma$--B for 1T\textquotesingle{}). Projections are shown as transparent red shaded regions. The minima and maxima for each band projection are shown as blue and red lines, respectively. The magnification highlights where band crossings occur in the T\textsubscript{d} phase.
    }
    \label{fig:ext-bulkbandprojections}
\end{extendeddatafigure}

\clearpage

\begin{extendeddatafigure}
    \centering
    \includegraphics[width=1\linewidth]{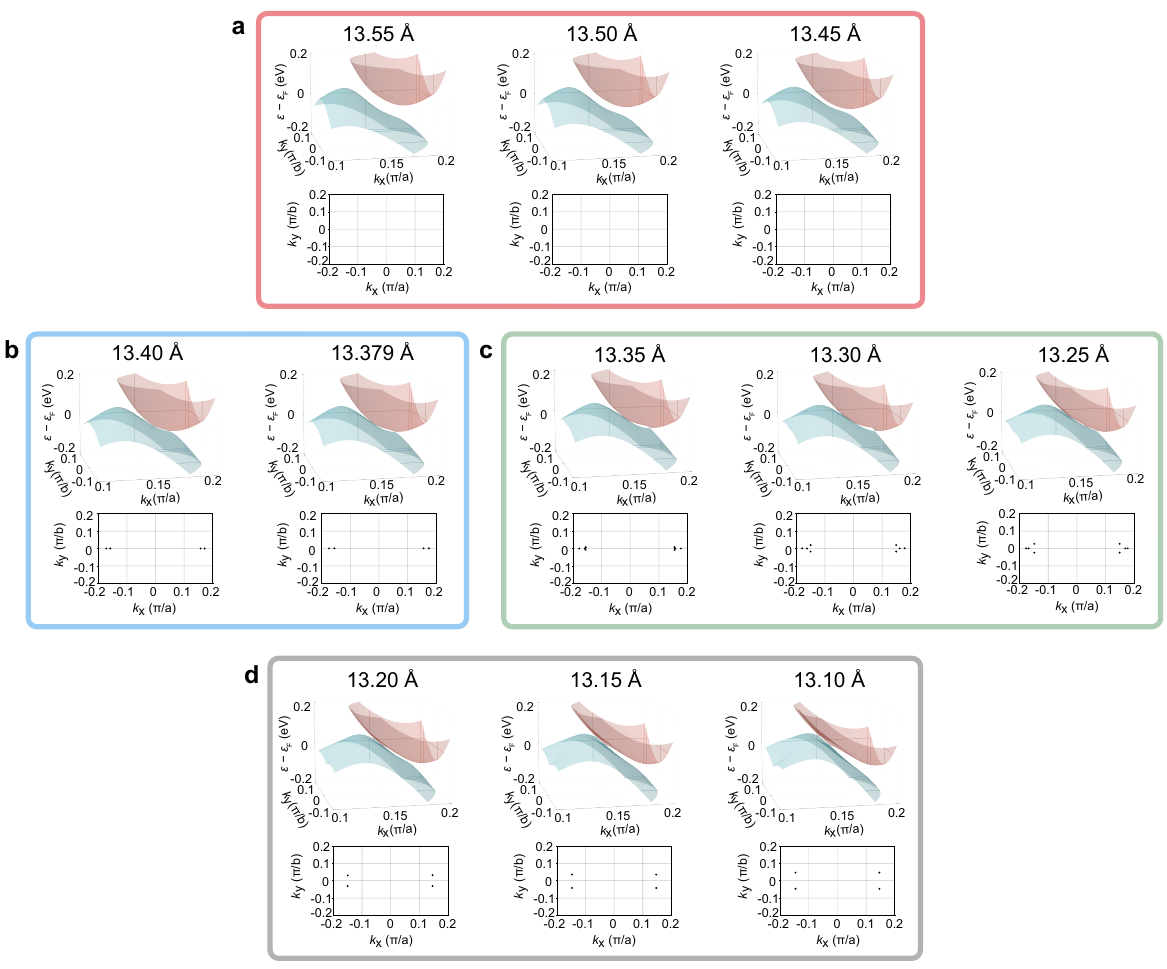}
    \caption{
    \textbf{Extended Data Figure 9 $\mid$ Calculated band structure for \ce{WTe2} with deformation strain along the $\vec{c}$-axis. } 
    \textbf{a}-\textbf{d}, Three-dimensional visualizations of the band structure for T\textsubscript{d}-\ce{WTe2} with compressive and tensile strain along the $\vec{c}$-axis (top) and the respective locations of their Weyl points in momentum space (bottom). The interplanar lattice constant, $c$, was fixed during the calculations at the value indicated above each band structure diagram. Atoms were allowed to relax within the constrained unit cells. The red (\textbf{a}), blue (\textbf{b}), green (\textbf{c}) and grey (\textbf{d}) enclosures identify deformation strains with zero, four, eight and four Weyl points within the first Brillouin zone, respectively. The Weyl points all lie at $k_z = 0$. The equilibrium structure has a lattice parameter of 13.379~\AA. An expansion in the $c$-parameter of 0.53\% is enough to make the Weyl points completely disappear (\textbf{a}), while a compression of only 0.22\% produces four additional Weyl points away from the high symmetry path, $\Gamma$--X (\textbf{c}). A further compression of 1.35\% relative to the equilibrium structure annihilates the four Weyl points along $\Gamma$--X, leaving four Weyl points located just off the high symmetry path (\textbf{d}). This extreme sensitivity of Weyl points to external strain is consistent with previous studies\cite{soluyanov2015type, Chang2016, DiSante2017, sie_ultrafast_2019, Rossi2020, Kim2017, Xu2017, Lv2017, Rubmann2018}.
    }
    \label{fig:ext-deformationstrain}
\end{extendeddatafigure}

\clearpage

\begin{extendeddatafigure}
    \centering
    \includegraphics[width=0.65\linewidth]{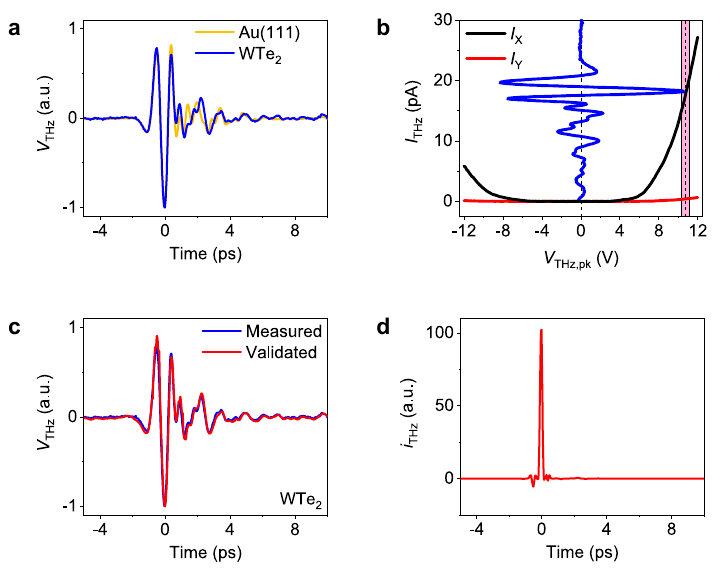}
    \caption{
    \textbf{Extended Data Figure 10 $\mid$ Waveform validation for THz-TDS in the \ce{WTe2} tunnel junction.}
    \textbf{a}, Terahertz cross-correlation (THz-CC) waveforms acquired on a \ce{WTe2} surface (blue curve) and Au(111) surface (gold curve). Parameters during waveform acquisition on \ce{WTe2}: $V_\mathrm{d.c.} = 1 V$, $I_\mathrm{d.c.} = \qty{100}{pA}$, $E_\mathrm{THz,pk} = \qty{+190}{V/cm}$, $E_\mathrm{WF,pk} = \qty{7}{V/cm}$, and on Au(111): $V_\mathrm{d.c.} = \qty{10}{mV}$, $I_\mathrm{d.c.} = \qty{100}{pA}$, $E_\mathrm{THz,pk} = \qty{+150}{V/cm}$, $E_\mathrm{WF,pk} = \qty{7}{V/cm}$. The data are shown as mean values of five individual scans. 
    \textbf{b}, $I_\mathrm{THz}-V_\mathrm{THz,pk}$ curve with both in-phase ($I_\mathrm{X}$, black curve) and out-of-phase ($I_\mathrm{Y}$, red curve) components of the terahertz-induced tunnel current ($I_\mathrm{THz}$). The strong terahertz pulse (blue curve) indicates where along the $I\textsubscript{THz}-V\textsubscript{THz,pk}$ curve waveform sampling takes place, while the pink shading shows the range of the weak terahertz pulse, with a peak field strength of $E_\mathrm{THz,pk} = \qty{+7}{V/cm}$ ($V_\mathrm{THz,pk} = \qty{+0.4}{V}$). 
    \textbf{c}, Measured (blue curve) and validated (red curve) THz-CC waveforms confirming that the waveform input to the simulation (blue curve) is an accurate representation of the terahertz voltage transient at the STM tip apex. 
    \textbf{d}, Simulated current pulse generated by the terahertz voltage waveform in \textbf{b} applied to the $I-V$ characteristic that was extracted from the $I_\mathrm{THz}-V_\mathrm{THz,pk}$ curve in \textbf{b} using a polynomial model\cite{Ammerman2022, Jelic_atomicTHzTDS_2024} with $N = 11$. The presence of a single unipolar current pulse validates the waveform measured on \ce{WTe2} via THz-CC (blue curve in \textbf{a}). A validation for the waveform on Au(111) that was used for this study is shown in ref.\,[\citen{Jelic_atomicTHzTDS_2024}].
    }
    \label{fig:ext-validation}
\end{extendeddatafigure}

\clearpage


\thispagestyle{empty}

\AddToShipoutPictureFG*{%
  \AtPageCenter{%
    \raisebox{-.5\height}{\makebox[0pt]{\fontsize{30}{20}\selectfont Supplementary Information}}
  }%
}
\mbox{}\addcontentsline{toc}{subsection}{Supplementary Information}

\clearpage

\setcounter{page}{1}

\clearpage

\begin{sifig}
    \centering
    \includegraphics[width=1\linewidth]{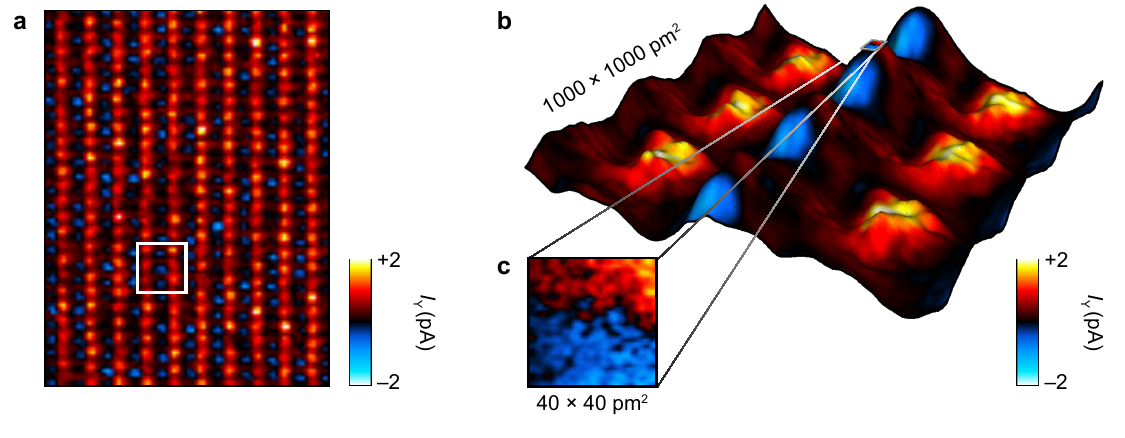}
    \caption{
    \textbf{Supplementary Figure 1 $\mid$ Picometre resolution tunnelling microscopy of a terahertz-driven phase transition in \ce{WTe2}.} 
    \textbf{a},~Out-of-phase THz-STM image, $I_\mathrm{Y}(x,y)$, acquired at $V\textsubscript{d.c.}$~=~10~mV, $I\textsubscript{d.c.}$~=~100~pA and $V\textsubscript{THz,pk}$~=~$-$1.7~V. Scan size 6~nm~$\times$~8~nm; scan speed 4.3~nm/s. \textbf{b}, Perspective view of $I_\mathrm{Y}(x,y)$ from Fig.\,\ref{fig:fig3}b overlaid onto a 3D texture generated by the simultaneously acquired $I\textsubscript{d.c.}(x,\,y)$, where $I\textsubscript{d.c.}$ spans 100~pA to 220~pA. \textbf{c}, Picometre-scale image from Fig.\,\ref{fig:fig3}d. The colourmap in \textbf{c} spans [$-1,\,+1$]~pA. The white box in \textbf{a} shows where \textbf{b} was acquired, while the grey box in \textbf{b} shows where \textbf{c} was acquired. 
    }
    \label{sifig:3D_THzSTM}
\end{sifig}

\clearpage

\begin{sifig}
    \centering
    \includegraphics[width=1\linewidth]{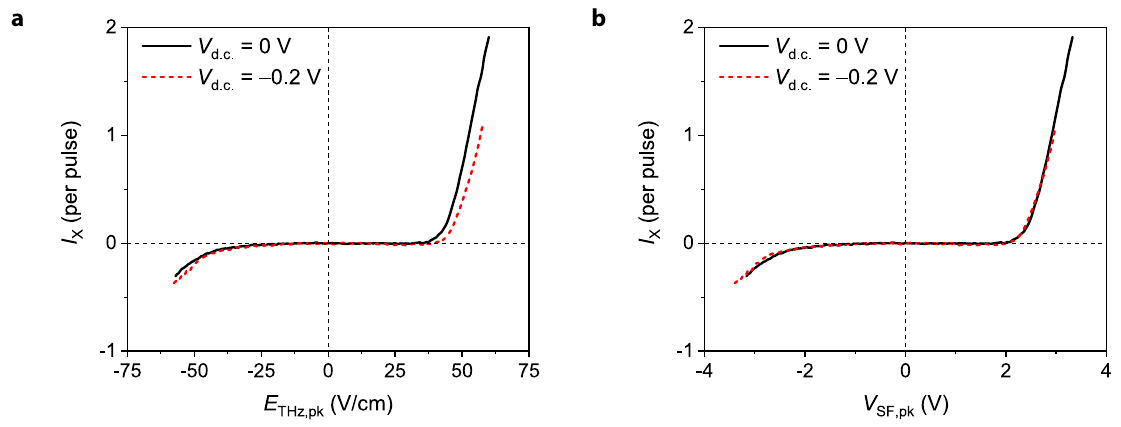}
    \caption{
    \textbf{Supplementary Figure 2 $\mid$ Terahertz pulse voltage calibration.} 
    \textbf{a},\textbf{b}, Voltage calibration performed for the terahertz waveform on \ce{WTe2}. A pair of $I_\mathrm{X}-E_\mathrm{THz,pk}$ curves (\textbf{a}) acquired at $V_\mathrm{d.c.} = \qty{0}{V}$ (solid black line) and $V_\mathrm{d.c.} = \qty{-0.2}{V}$ (dashed red line) are used to calibrate the peak terahertz voltage\cite{Jelic_atomicTHzTDS_2024} by translating the dashed red line along the x-axis by \qty{3.6}{V/cm} (\textbf{b}), equivalent to \qty{0.20}{V}, which results in a calibration constant of $\alpha = (\qty{1}{V})/(\qty{18}{V/cm})$. The measurements were performed on \ce{WTe2} at constant tip height after a \qty{200}{pm} tip retraction from an initial tip–sample separation set by $V\textsubscript{d.c.} = \qty{10}{mV}$, $I\textsubscript{d.c.} = \qty{100}{pA}$.
    }
    \label{sifig:thzvoltagecalibration}
\end{sifig}

\clearpage

\begin{sifig}
    \centering
    \includegraphics[width=1\linewidth]{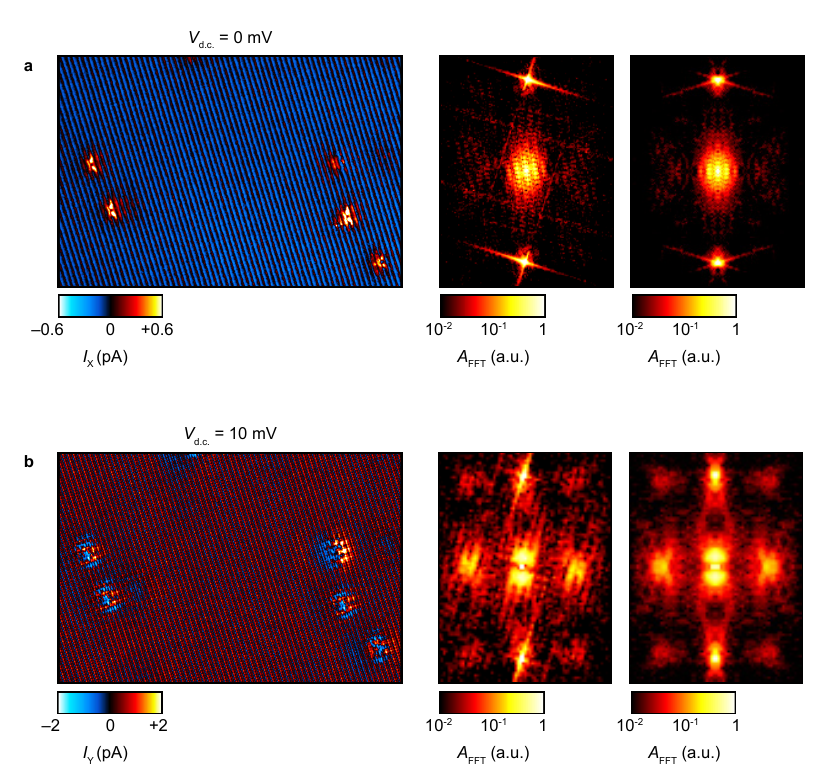}
    \caption{
    \textbf{Supplementary Figure 3 $\mid$ Large-area differential atomic imaging of a terahertz-field-driven phase transition.} 
    \textbf{a},~Conventional in-phase THz-STM rectified current map, $I\textsubscript{X}(x,\,y)$ (left), and symmetrized 2D FFT, $I\textsubscript{X}(k\textsubscript{x},\,k\textsubscript{y})$ (right), from Fig.\,\ref{fig:fig2}d with the non-symmetrized 2D FFT shown in the middle.
    \textbf{b},~Out-of-phase THz-STM image (left) from Fig.\,\ref{fig:fig2}e, showing $I_\mathrm{Y}(x,y)$ for the same area in \textbf{a}. The symmetrized 2D FFT, $I\textsubscript{Y}(k\textsubscript{x},\,k\textsubscript{y})$ (right), is shown beside the non-symmetrized 2D FFT (middle). The cross-hatch features surrounding bright peaks in the momentum space images are inevitable artifacts of the windowless 2D FFTs that were performed.
    }
    \label{sifig:THzSTMqpi}
\end{sifig}

\clearpage


\begin{table}
    \centering
    \begin{tabular}{|c|c|}
        \hline
        Frequency $(\text{cm}^{\text{-1}}\rule{0pt}{10pt})$ & Frequency (THz) \\ \hline
        9.6821 &   0.2903   \\
        24.6427 &   0.7388 \\
        28.7741  &  0.8626 \\
        84.9640  &  2.5472 \\
        94.6442  &  2.8374 \\
        101.0263  &  3.0287 \\
        101.9236   & 3.0556 \\
        125.2583  &  3.7551 \\
        127.1769  &  3.8127 \\
        131.6546  &  3.9469 \\ \hline
    \end{tabular}
    \caption{
    \textbf{Supplementary Table 1 $\mid$ Calculated phonon frequencies in T\textsubscript{d}-\ce{WTe2} at the $\Gamma$ point.} Vibrational frequencies calculated for a two-layer slab using a 2$\times$2 supercell.
    }
    \label{tab:sitab-phonondispersion}
\end{table}

\clearpage

\begin{sidewaystable}
\centering
\fontsize{7.2}{10}\selectfont
\begin{tabular}{|c|c|c|c|c|c|c|c|c|c|c|c|c|c|}
\hline
& & & & & & & & \multicolumn{3}{|c|}{\textbf{Lattice Constants}} & & & \\ \hline
\textbf{Article} & \textbf{\makecell{Simulation \\ Package}} & \textbf{\makecell{Functional \\ Class}} & \textbf{Functional} & \textbf{\makecell{DFT Basis Set \\ (Energy Cutoff)}} & \textbf{\makecell{Spin-Orbit \\ Coupling}} & \textbf{\makecell{Dispersion \\ Correction}} & \textbf{\textit{a} (\AA)} & \textbf{\textit{b} (\AA)} & \textbf{\textit{c} (\AA)} & \textbf{\makecell{Unit Cell \\ Volume \\(\AA$^3$)}} & \textbf{\makecell{Relaxed \\ Lattice?}} & \textbf{\makecell{Weyl \\ Points}} & \textbf{Notes}\\ \hline
Soluyanov [\citen{soluyanov2015type}]& VASP & GGA & PBE & PW+PAW (260 eV) & SOC &not included& Exp.\cite{mar_Tdstructure_1992} & - & - & 305\cite{mar_Tdstructure_1992} & no & 8 WPs\textsuperscript{a} & a \\ \hline
Chang [\citen{Chang2016}]& VASP & GGA & PBE & PW+PAW (n.s.) & SOC &not included& 3.453 & 6.328 & 13.506 & 295 & yes & 0 WPs\textsuperscript{b} & b \\ \hline
Bruno [\citen{Bruno_observ_2016}]& VASP & GGA & PBE & PW+PAW (n.s.) & SOC &not included& Exp.\cite{mar_Tdstructure_1992} & - & - & 305\cite{mar_Tdstructure_1992} & no & 8 WPs (<\,113 K) & \\ \hline
Wang [\citen{Wang_observation_2016}]& VASP & GGA & PBE & PW+PAW (500 eV) & SOC &not included& Exp.\cite{mar_Tdstructure_1992} & - & - & 305\cite{mar_Tdstructure_1992} & no & 8 WPs (<\,113 K) & \\ \hline
Di Sante [\citen{DiSante2017}]& VASP \& GW & LDA & LDA+U & PW+PAW (400 eV) & SOC &not included& n.s. & n.s. & n.s. & n.s. & yes & $\geq$\,4 WPs\textsuperscript{c} & c \\ \hline
Crepaldi [\citen{Crepaldi2017}]& QE & GGA & PBE & PW+NCP (n.s.) & SOC &not included& Exp.\cite{mar_Tdstructure_1992, Brown1966} & - & - & 305\cite{mar_Tdstructure_1992} & no & 0 WPs\textsuperscript{d} & d \\ \hline
Zhang [\citen{Zhang2017RRL}]& VASP & GGA & PBE & PW+PAW (300 eV) & SOC &not included& 3.456 & 6.213 & 13.935 & 299 & yes & 8 WPs (<\,10 K) & \\ \hline
Kim [\citen{Kim2017}]& \makecell{ VASP \\ VASP \\ VASP \\ VASP \\ VASP \\ VASP \\ VASP \\ VASP } & \makecell{ LDA \\ GGA \\ GGA \\ GGA \\ GGA \\ GGA \\ GGA \\ GGA } & \makecell{ LDA \\ PBE \\ PBE \\ PBE \\ vdW-DF \\ vdW-DF2 \\ optB88-vdW \\ rev-vdW-DF2 } & \makecell{ PW+PAW (450 eV) \\ PW+PAW (450 eV) \\ PW+PAW (450 eV) \\ PW+PAW (450 eV) \\ PW+PAW (450 eV) \\ PW+PAW (450 eV) \\ PW+PAW (450 eV) \\ PW+PAW (450 eV) } & \makecell{ SOC \\ SOC \\ SOC \\ SOC \\ SOC \\ SOC \\ SOC \\ SOC } & \makecell{not included\\not included\\ D2 \\ D3 \\ functional \\ functional \\ functional \\ functional } & \makecell{ 3.454 \\ 3.507 \\ 3.544 \\ 3.479 \\ 3.597 \\ 3.681 \\ 3.545 \\ 3.509 } & \makecell{ 6.208 \\ 6.311 \\ 6.225 \\ 6.279 \\ 6.392 \\ 6.484 \\ 6.309 \\ 6.276 } & \makecell{ 13.818 \\ 15.698 \\ 13.761 \\ 14.042 \\ 15.162 \\ 15.023 \\ 14.363 \\ 14.171 } & \makecell{ 296 \\ 347 \\ 304 \\ 307 \\ 349 \\ 359 \\ 321 \\ 312 } & \makecell{ yes \\ yes \\ yes \\ yes \\ yes \\ yes \\ yes \\ yes } & \makecell{ 0 WPs \\ 0 WPs \\ 0 WPs \\ 0 WPs \\ 0 WPs \\ 0 WPs \\ 0 WPs \\ 0 WPs\textsuperscript{e} } & \makecell{   \\   \\   \\   \\   \\   \\   \\ e } \\ \hline
Lin [\citen{lin_visualizing_2017}]& VASP & GGA & PBE & PW+PAW (400 eV) & SOC &not included& Exp.\cite{mar_Tdstructure_1992} & - & - & 305\cite{mar_Tdstructure_1992} & no & 8 WPs & \\ \hline
Xu [\citen{Xu2017}]& \makecell{ VASP \\ VASP } & \makecell{ GGA \\ GGA } & \makecell{ PBE \\ PBE } & \makecell{ PW+PAW (320 eV) \\ PW+PAW (320 eV) } & \makecell{ SOC \\ SOC } & \makecell{not included\\not included} & \makecell{ n.s. \\ n.s. } & \makecell{ n.s. \\ n.s. } & \makecell{ 14.15 \\ 14.01 } & \makecell{ 307 \\ 304 } & \makecell{ no \\ no } & \makecell{ 0 WPs \\ 8 WPs } & \\ \hline
Lv [\citen{Lv2017}]\textsuperscript{f}& \makecell{ VASP \\ VASP \\ VASP \\ VASP \\ VASP} & \makecell{ GGA \\ GGA \\ GGA \\ GGA \\ GGA} & \makecell{ PBE \\ PBE \\ PBE \\ PBE \\ PBE} & \makecell{ PW+PAW (300 eV) \\ PW+PAW (300 eV) \\ PW+PAW (300 eV) \\ PW+PAW (300 eV) \\ PW+PAW (300 eV)} & \makecell{ SOC\textsuperscript{g} \\ SOC\textsuperscript{g} \\ SOC\textsuperscript{g} \\ SOC\textsuperscript{g} \\ SOC\textsuperscript{g}} & \makecell{not included\\not included\\not included\\not included\\not included} & \makecell{ 3.463\textsuperscript{h} \\ 3.467\textsuperscript{h} \\ 3.470\textsuperscript{h} \\ 3.482\textsuperscript{h} \\ 3.489\textsuperscript{h}} & \makecell{ 6.230\textsuperscript{h} \\ 6.230\textsuperscript{h} \\ 6.238\textsuperscript{h} \\ 6.256\textsuperscript{h} \\ 6.266\textsuperscript{h}} & \makecell{ 14.004\textsuperscript{h} \\ 14.017\textsuperscript{h} \\ 14.019\textsuperscript{h} \\ 14.059\textsuperscript{h} \\ 14.083\textsuperscript{h}} & \makecell{ 302 \\ 303 \\ 303 \\ 306 \\ 308} & \makecell{ yes\textsuperscript{g} \\ yes\textsuperscript{g} \\ yes\textsuperscript{g} \\ yes\textsuperscript{g} \\ yes\textsuperscript{g}} & \makecell{ 8 WPs (35 K)\textsuperscript{f} \\ 8 WPs (60 K)\textsuperscript{f} \\ 0 WPs (100 K)\textsuperscript{f} \\ 0 WPs (200 K)\textsuperscript{f} \\ 0 WPs (300 K)\textsuperscript{f}} & \makecell{ f--h \\ f--h \\ f--h \\ f--h \\ f--h}\\ \hline
Yuan [\citen{Yuan_qpi_2018}]& VASP & GGA & PBE & PW+PAW (300 eV) &not included&not included& n.s. & n.s. & n.s. & n.s. & no & 8 WPs & \\ \hline
R\"u\ss mann [\citen{Rubmann2018}]& KKR & LDA & LSDA & FLAPW (n.s.) & SOC &not included& Exp.\cite{mar_Tdstructure_1992} & - & - & 305\cite{mar_Tdstructure_1992} & no & 0 WPs\textsuperscript{i} & i \\ \hline
Sie [\citen{sie_ultrafast_2019}]&  \makecell{ VASP \\ VASP \\ VASP} & \makecell{ GGA \\ GGA \\ GGA} & \makecell{ PBE \\ PBE \\ PBE} & \makecell{ PW+PAW (260 eV) \\ PW+PAW (400 eV) \\ PW+PAW (400 eV)} & \makecell{ SOC \\ SOC \\ SOC} & \makecell{not included\\not included\\ D3-BJ} & \makecell{ Exp.\cite{mar_Tdstructure_1992} \\ 3.481 \\ 3.484} & \makecell{ - \\ 6.281 \\ 6.253} & \makecell{ - \\ 13.997 \\ 13.591} & \makecell{ 305\cite{mar_Tdstructure_1992} \\ 306 \\ 296} & \makecell{ no \\ yes \\ yes} & \makecell{ 8 WPs \\ 8 WPs \\ 8 WPs} & \\ \hline
Rossi [\citen{Rossi2020}]& OpenMX & GGA & PBE & PAO+NCP (2.7 keV) & SOC &not included& Exp.\cite{mar_Tdstructure_1992, Brown1966} & - & - & 305\cite{mar_Tdstructure_1992} & no & 0 WPs\textsuperscript{j} & j \\ \hline
Das [\citen{Das_electronicproperties_2019}]& VASP &GGA \& LDA&PBE \& LDA+U& PW+PAW (n.s.) & SOC &not included& n.s. & n.s. & n.s. & n.s. & yes & 8 WPs & \\ \hline
Hein [\citen{Hein2020mode}]& VASP & GGA & PBE & PW+PAW (350 eV) & SOC & D2 & Exp.\cite{mar_Tdstructure_1992} & - & - & 305\cite{mar_Tdstructure_1992} & no & 8 WPs & \\ \hline
Guan [\citen{guan_manipulating_2021}]& VASP & GGA & PBE & PW+PAW (300 eV) &not included&not included& Exp.\cite{mar_Tdstructure_1992} & - & - & 305\cite{mar_Tdstructure_1992} & no & 8 WPs & \\ \hline
this article& \makecell{ CRYSTAL \\ CRYSTAL \\ CRYSTAL}  & \makecell{ GGA \\ GGA \\ hybrid-GGA} & \makecell{ PBE \\ PBE \\ PBE0 } & \makecell{ GTO \\ GTO \\ GTO} & \makecell{ SOC\textsuperscript{g} \\ SOC\textsuperscript{g} \\ SOC\textsuperscript{k}} & \makecell{not included\\ D3-BJ \\ D3-BJ} & \makecell{ 3.472 \\ 3.448 \\ 3.408} & \makecell{ 6.252 \\ 6.215 \\ 6.135} & \makecell{ 14.374 \\ 13.216 \\ 13.380} & \makecell{ 312 \\ 283 \\ 280} & \makecell{ yes\textsuperscript{g} \\ yes\textsuperscript{g} \\ yes\textsuperscript{k}} & \makecell{ 0 WPs \\ 0 WPs \\ 4 WPs\textsuperscript{l}} & \makecell{ g \\ g \\ k,l}\\ \hline

\end{tabular}
\captionsetup{font=scriptsize}
\caption{\textbf{Supplementary Table 2 | Summary of lattice parameters and Weyl points in bulk T\textsubscript{d}-\ce{WTe2} explored with DFT.} Most of the articles emphasize that the presence of Weyl points is highly sensitive to external parameters such as strain, doping, and temperature. Several studies do not relax the atomic lattice positions within DFT and instead use the lattice constants determined by X-ray diffraction, which is denoted by \textquotesingle{}Exp.\textquotesingle{} within the first Lattice Constants column (\textit{a}~=~3.477~\AA, \textit{b}~=~6.249~\AA, \textit{c} = 14.018 \AA)\cite{mar_Tdstructure_1992}. These lattice constants are in agreement with other X-ray diffraction studies\cite{Brixner1962, Brown1966, Obolonchik1972, Pan2015, Lv2017}. DFT - density functional theory; GW – computational technique utilizing Green’s functions and a screened Coulomb interaction; VASP – Vienna ab initio simulation package; QE - Quantum ESSPRESSO simulation package; KKR – Korringa-Kohn-Rostoker Green's function method; GGA – generalized gradient approximation; PW - plane wave; GTO - Gaussian-type orbital; PAW – projector augmented-wave; NCP – norm-conserving pseudopotential; PAO – pseudo-atomic orbital; PBE – Perdew-Burke-Ernzerhof functional; PBE0 – hybrid PBE functional with 75\% exchange interaction from PBE and 25\% from Hartree-Fock; FLAPW – full-potential linearized augmented plane wave method; LDA – local density approximation functional; LSDA – local spin density approximation functional; LDA+U – LDA with a Hubbard correction for on-site Coulomb interactions; vdW-DF – van der Waals density functional; vdW-DF2 – van der Waals density functional version \#2; rev-vdW-DF2 – revised van der Waals density functional version \#2; optB88-vdW – optimized Becke88 van der Waals functional; SOC – spin-orbit coupling; D3-BJ – D3 dispersion correction with Becke-Johnson damping; WPs – Weyl points; n.s. – parameter was not specified.}
\caption*{\fontsize{6}{7}\selectfont
    \parbox[t]{0.45\textwidth}{\raggedright a: Without SOC included, the calculations find 16 Weyl points. \\ b: 0.9 meV gap (no Weyl points). \\ c: At least four Weyl points are present according to Figure S5, but no value was explicitly stated. \\ d: 8 meV gap (no Weyl points). However, Weyl points emerge when using VASP instead of Quantum ESPRESSO. \\ e: Eight Weyl points are present with a biaxial strain of –1.5\% along $\vec{a}$ \& $\vec{c}$. \\ f: Calculations are explicitly for WTe\textsubscript{1.98}.
}%
    \hfill
    \parbox[t]{0.45\textwidth}{\raggedright g: Atomic lattice relaxation is performed without SOC using PBE, but SOC is included in the band structure calculations. \\ h: These are powder X-ray diffraction experimental lattice constants. Atomic lattice positions after relaxation are not reported. \\ i: 15 meV gap (no Weyl points). \\ j: Eight Weyl points are present with a compressive strain of $\sim$1.5\% along $\vec{c}$ \\ k: Atomic lattice relaxation is performed without SOC using PBE0, but SOC is included in the band structure calculations. \\ l: Eight Weyl Points are present with a compressive strain of 0.22\% along $\vec{c}$.}
}
\label{sitab:Td_refs}
\end{sidewaystable}

\clearpage

\begin{sidewaystable}
\centering
\fontsize{7.7}{10}\selectfont
\begin{tabular}{|c|c|c|c|c|c|c|c|c|c|c|c|c|}
\hline
& & & & & & & & \multicolumn{3}{|c|}{\textbf{Lattice Constants}} & & \\ \hline
\textbf{Article} & \textbf{\makecell{Simulation \\ Package}} & \textbf{\makecell{Functional \\ Class}} & \textbf{Functional} & \textbf{\makecell{DFT Basis Set \\ (Energy Cutoff)}} & \textbf{\makecell{Spin-Orbit \\ Coupling}} & \textbf{\makecell{Dispersion \\ Correction}} & \textbf{\textit{a} (\AA)} & \textbf{\textit{b} (\AA)} & \textbf{\textit{c} (\AA)} & \textbf{\makecell{Unit Cell \\ Volume \\(\AA$^3$)}} & \textbf{\makecell{Relaxed \\ Lattice?}} & \textbf{\makecell{Weyl \\ Points}} \\ \hline
Ali [\citen{Ali2014}]& WIEN2k & GGA & PBE & LAPW ($R\textsubscript{mt} \times K\textsubscript{max}$ = 8) & SOC &not included& Exp.\cite{Brown1966} & - & - & 309\cite{Brown1966} & no & n.s.\\ \hline
Jiang [\citen{Jiang2015}]& WIEN2k & GGA & PBE & LAPW ($R\textsubscript{mt} \times K\textsubscript{max}$ = 7) & SOC &not included& n.s. & n.s. & n.s. & n.s. & no & n.s. \\ \hline
Sánchez-Barriga [\citen{SanchezBarriga2016}]& VASP & GGA & PBEsol & PW+PAW (n.s.) &not included&not included& Exp.\cite{Brown1966} & - & - & 309\cite{Brown1966} & no & n.s. \\ \hline
Jiang [\citen{Jiang2016}]& CASTEP & LDA & LDA & PW+PAW (500 eV) & SOC &not included& Exp.\cite{Brown1966} & - & - & 309\cite{Brown1966} & no & n.s. \\ \hline
Feng [\citen{Feng_spin_2016}]& VASP & GGA & PBE & PW+PAW (360 eV) & SOC &not included& Exp.\cite{mar_Tdstructure_1992} & - & - & 305\cite{mar_Tdstructure_1992} & no & n.s. \\ \hline
Song [\citen{Song2016}]& VASP & LDA & LDA & PW+PAW (350 eV) & SOC &not included& Exp.\cite{mar_Tdstructure_1992} & - & - & 305\cite{mar_Tdstructure_1992} & no & n.s. \\ \hline
Zhang [\citen{Zhang_qpi_2017}]& VASP & GGA & PBE & PW+PAW (450 eV) & SOC &not included& n.s. & n.s. & n.s. & n.s. & no & n.s. \\ \hline
Sharma [\citen{Sharma2019}]&VASP \& QE& GGA & PBE & PW+PAW (n.s.) & SOC &not included& n.s. & n.s. & n.s. & n.s. & yes & n.s. \\ \hline
Ni [\citen{Ni2020}]& VASP & GGA &optB86b-vdW& PW+PAW (400 eV) & SOC & functional & 3.49 & 6.28 & 14.24 & 312 & yes & n.s. \\ \hline
Hein [\citen{Hein2020JPCM}]& VASP & GGA & PBE & PW+PAW (350 eV) & SOC &not included& Exp.\cite{mar_Tdstructure_1992} & - & - & 305\cite{mar_Tdstructure_1992} & no & n.s. \\ \hline
Ji [\citen{ji_manipulation_2021}]& VASP & GGA & PBE & PW+PAW (446 eV) & SOC & D3-BJ & n.s. & n.s. & n.s. & n.s. & yes & n.s. \\ \hline
Soranzio [\citen{Soranzio2022}]& QE & GGA & PBE & PW+PAW (1088 eV) & SOC &not included& n.s. & n.s. & n.s. & n.s. & yes & n.s. \\ \hline
Chen [\citen{Chen_noncentrosym_2022}]& VASP & GGA & PBE & PW+PAW (400 eV) & SOC &not included& n.s. & n.s. & n.s. & n.s. & no & n.s. \\ \hline

\end{tabular}
\captionsetup{font=scriptsize}
\caption{\textbf{Supplementary Table 3 | Investigating bulk T\textsubscript{d}-\ce{WTe2} with DFT.} Most studies do not relax the atomic lattice positions within DFT and instead use lattice constants determined by experiment\cite{mar_Tdstructure_1992, Brixner1962, Brown1966, Obolonchik1972, Pan2015, Lv2017}, which is denoted by \textquotesingle{}Exp.\textquotesingle{} within the first Lattice Constants column. DFT - density functional theory; VASP – Vienna ab initio simulation package; QE - Quantum ESSPRESSO simulation package; CASTEP – Cambridge serial total energy package; WIEN2k – DFT simulation package; GGA – generalized gradient approximation; PW - plane wave; PAW – projector augmented-wave; PBE – Perdew-Burke-Ernzerhof functional; PBEsol – PBE functional optimized for crystalline solids; LAPW – linearized augmented plane wave method; LDA – local density approximation functional; optB86b-vdW – optimized Becke86b van der Waals functional; SOC – spin-orbit coupling; D3-BJ – D3 dispersion correction with Becke-Johnson damping; WPs – Weyl points; n.s. – parameter was not specified.}
\label{sitab:Td_refs2}
\end{sidewaystable}

\clearpage

\begin{sidewaystable}
\centering
\fontsize{7.7}{10}\selectfont
\begin{tabular}{|c|c|c|c|c|c|c|c|c|c|c|c|c|c|c|}
\hline
& & & & & & & & \multicolumn{3}{|c|}{\textbf{Lattice Constants}} & & & & \\ \hline
\textbf{Article} & \textbf{\makecell{Simulation \\ Package}} & \textbf{\makecell{Functional \\ Class}} & \textbf{Functional} & \textbf{\makecell{DFT Basis Set \\ (Energy Cutoff)}} & \textbf{\makecell{Spin-Orbit \\ Coupling}} & \textbf{\makecell{Dispersion \\ Correction}} & \textbf{\textit{a} (\AA)} & \textbf{\textit{b} (\AA)} & \textbf{\textit{c} (\AA)} & \textbf{\makecell{Monoclinic \\ Angle ($^{\circ}$) }} & \textbf{\makecell{Unit Cell \\ Volume \\(\AA$^3$)}} & \textbf{\makecell{Relaxed \\ Lattice?}} & \textbf{\makecell{Weyl \\ Points}} & \textbf{Notes}\\ \hline
Kim [\citen{Kim2017}]& \makecell{ VASP \\ VASP \\ VASP \\ VASP \\ VASP \\ VASP \\ VASP \\ VASP } & \makecell{ LDA \\ GGA \\ GGA \\ GGA \\ GGA \\ GGA \\ GGA \\ GGA } & \makecell{ LDA \\ PBE \\ PBE \\ PBE \\ vdW-DF \\ vdW-DF2 \\ optB88-vdW \\ rev-vdW-DF2 } & \makecell{ PW+PAW (450 eV) \\ PW+PAW (450 eV) \\ PW+PAW (450 eV) \\ PW+PAW (450 eV) \\ PW+PAW (450 eV) \\ PW+PAW (450 eV) \\ PW+PAW (450 eV) \\ PW+PAW (450 eV) } & \makecell{ SOC \\ SOC \\ SOC \\ SOC \\ SOC \\ SOC \\ SOC \\ SOC } & \makecell{not included\\not included\\ D2 \\ D3 \\ functional \\ functional \\ functional \\ functional } & \makecell{ 3.453 \\ 3.507 \\ 3.543 \\ 3.479 \\ 3.597 \\ 3.681 \\ 3.545 \\ 3.508 } & \makecell{ 6.210 \\ 6.312 \\ 6.227 \\ 6.282 \\ 6.392 \\ 6.484 \\ 6.310 \\ 6.278 } & \makecell{ 13.810 \\ 15.647 \\ 13.764 \\ 14.029 \\ 15.188 \\ 15.042 \\ 14.363 \\ 14.173 } & \makecell{ 92.2 \\ 91.4 \\ 91.1 \\ 92.0 \\ 90.8 \\ 91.3 \\ 90.6 \\ 92.1 } & \makecell{ 296 \\ 346 \\ 304 \\ 306 \\ 349 \\ 359 \\ 321 \\ 312 } & \makecell{ yes \\ yes \\ yes \\ yes \\ yes \\ yes \\ yes \\ yes } & \makecell{ 0 WPs \\ 0 WPs \\ 0 WPs \\ 0 WPs \\ 0 WPs \\ 0 WPs \\ 0 WPs \\ 0 WPs } & \\ \hline
Rossi [\citen{Rossi2020}]& OpenMX & GGA & PBE & PAO+NCP (2.7 keV) & SOC &not included& Exp.\cite{Brown1966} & - & - & 93.6\cite{Brown1966} & 304\cite{Brown1966} & no\textsuperscript{a} & 0 WPs & a \\ \hline
this article& \makecell{ CRYSTAL \\ CRYSTAL \\ CRYSTAL}  & \makecell{ GGA \\ GGA \\ hybrid-GGA} & \makecell{ PBE \\ PBE \\ PBE0 } & \makecell{ GTO \\ GTO \\ GTO} & \makecell{ SOC\textsuperscript{b} \\ SOC\textsuperscript{b} \\ SOC\textsuperscript{c}} & \makecell{not included\\ D3-BJ \\ D3-BJ} & \makecell{ 3.469 \\ 3.454 \\ 3.406} & \makecell{ 6.249 \\ 6.210 \\ 6.133} & \makecell{ 14.419 \\ 13.195 \\ 13.448} & \makecell{ 89.5 \\ 93.9 \\ 94.2} & \makecell{ 313 \\ 282 \\ 280} & \makecell{ yes\textsuperscript{b} \\ yes\textsuperscript{b} \\ yes\textsuperscript{c}} & \makecell{ 0 WPs \\ 0 WPs \\ 0 WPs} & \makecell{ b \\ b \\ c}\\ \hline

\end{tabular}
\captionsetup{font=scriptsize}
\caption{\textbf{Supplementary Table 4 | Summary of lattice parameters in bulk 1T\textquotesingle{}-\ce{WTe2} explored with DFT.} Since experimental values for the lattice constants are not readily available for \ce{WTe2}, the initial lattice constants before atomic relaxation were set by the relatively similar crystal structure of monoclinic 1T\textquotesingle{}-\ce{MoTe2} (\textit{\textbf{a}}~=~3.469 \AA, \textit{\textbf{b}} = 6.33 \AA, \textit{\textbf{c}} = 13.86 \AA, \textit{\textbf{$\theta$}} = 93.55$^\circ$)\cite{Brown1966}. DFT - density functional theory; VASP – Vienna ab initio simulation package; GGA – generalized gradient approximation; PW - plane wave; GTO - Gaussian-type orbital; PAW – projector augmented-wave; NCP – norm-conserving pseudopotential; PAO – pseudo-atomic orbital; PBE – Perdew-Burke-Ernzerhof functional; PBE0 – hybrid PBE functional with 75\% exchange interaction from PBE and 25\% from Hartree-Fock; LDA – local density approximation functional; vdW-DF – van der Waals density functional; vdW-DF2 – van der Waals density functional version \#2; rev-vdW-DF2 – revised van der Waals density functional version \#2; optB88-vdW – optimized Becke88 van der Waals functional; SOC – spin-orbit coupling; D3-BJ – D3 dispersion correction with Becke-Johnson damping; WPs – Weyl points.}
\caption*{\fontsize{6}{7}\selectfont
    \parbox[t]{0.45\textwidth}{\raggedright a: Atomic lattice relaxation is not performed. Instead, the experimental values for \ce{MoTe2} are used [\citen{Brown1966}]. \\ b: Atomic lattice relaxation is performed without SOC using PBE, but SOC is included in the band structure calculations. \\ c: Atomic lattice relaxation is performed without SOC using PBE0, but SOC is included in the band structure calculations.
}%
    \hfill
    \parbox[t]{0.45\textwidth}{\raggedright}
}
\label{sitab:1Tp_refs}
\end{sidewaystable}

\end{document}